\documentclass{emulateapj}

\usepackage{pslatex}

\newcommand{\vpp}{$v\arcsec$}
\newcommand{\Jpp}{$J\arcsec$}

\newcommand{\Htwo}{\mbox{H$_{2}$}}

\newcommand{\kms}{km~s$^{-1}$}

\newcommand{\ergs}{ergs cm$^{-2}$ s$^{-1}$ \AA$^{-1}$}
\newcommand{\lam}{$\lambda$}
\newcommand{\dlam}{$\lambda\lambda$}
\newcommand{\HST}{{\em HST}}

\newcommand{\FUSE}{{\em FUSE}}
\newcommand{\lya}{{Ly$\alpha$}}

\journalinfo{ACCEPTED FOR PUBLICATION IN THE ASTROPHYSICAL JOURNAL SUPPLEMENT SERIES}
\submitted{Received 2006 November 10; accepted 2006 December 14 }

\shorttitle{Metal Absorption Profiles from M27}
\shortauthors{McCandliss and Kruk}

\begin{document}

\title{Metal Absorption Profiles from the Central Star of the Planetary Nebula
M27 (NGC 6853, PN G060.8-03.6, the Dumbbell) -- Photospheric and Nebular
Line Identifications
}


\author{Stephan. R. McCandliss\altaffilmark{1} \& Jeffrey Kruk\altaffilmark{1}}
\affil{Department of Physics and Astronomy,
The Johns Hopkins University,
Baltimore, MD  21218.}
\email{stephan@pha.jhu.edu}






\begin{abstract}

High resolution spectra of the hot central star of the planetary nebula
(CSPN) M27, acquired with the Far Ultraviolet Spectroscopic Explorer
(\FUSE), have revealed an unusually rich set of narrow molecular
hydrogen absorption features.  This object is also unique in that the
velocity of nebular absorption features are cleanly separated from the
velocity of the intervening interstellar medium.  These features blend with
and in many cases obscure atomic features.  We have developed a
continuum model of the CSPN including atomic and molecular hydrogen
absorption.  Using this model we have identified and tabulated the
metal lines as arising from either photospheric, nebular
and/or non-nebular velocity systems.  We find the nebular outflow and
ionization balance to be stratified with high ionization states favored
at low velocity and low ionization states favored at high velocity.
Neutrals and molecules are found at a velocity that marks the
transistion between these two regimes.  These observations are a
challenge to the interacting wind model of PN evolution. Mappings at
high resolution of the line profiles for \ion{C}{1} \ion{-}{4},
\ion{N}{1} \ion{-}{3}, \ion{O}{1}, \ion{O}{6}, \ion{Si}{2} \ion{-}{4},
\ion{P}{2} \ion{-}{5}, \ion{S}{2} \ion{-}{4}, \ion{Ar}{1} \ion{-}{2}
and \ion{Fe}{2} \ion{-}{3} within the \FUSE\ and STIS bandpasses are
presented.  The digitial spectra of the star and the model are freely
available on the H$_2$ools website.  They will be useful for
photospheric analyses seeking to determine the metallicity of the
central star and for absorption line based atomic and molecular
abundance determination of the nebular outflow.

\end{abstract}





\section{Introduction}

\begin{figure*}[t]
\includegraphics[]{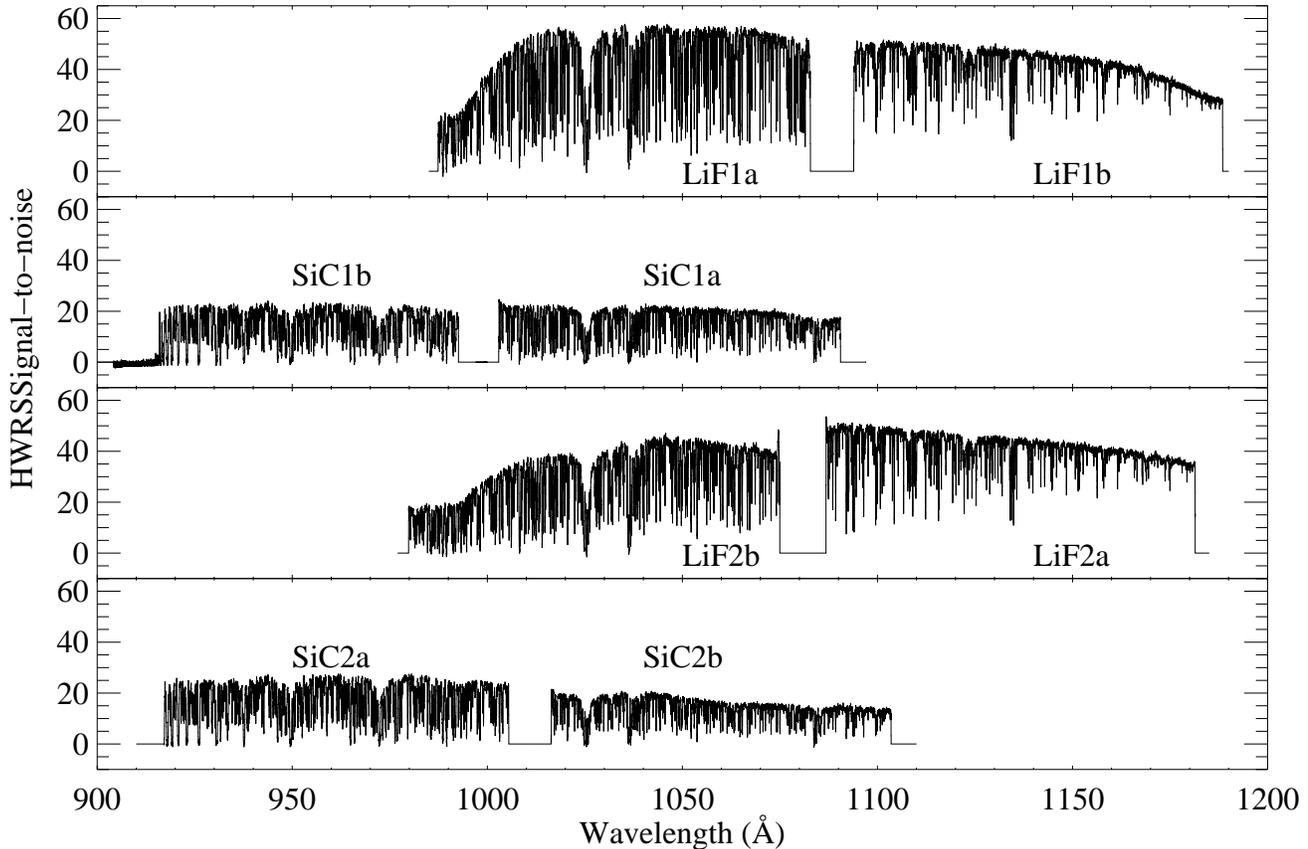}
\vspace*{.35in}
\caption[f1.eps]{ \label{signos} 
Signal-to-noise ratio of the eight coadded spectral segments taken
through the HIRS spectrograph aperture.  The individual segments are
labeled. 
}
\end{figure*}

High resolution spectra of the hot central star (CS) of the Planetary Nebula (PN) M27 (NGC~6853, PN G060.8-03.6, the Dumbbell), acquired with the Far Ultraviolet Spectroscopic Explorer
(\FUSE), have revealed an unusually rich set of narrow molecular
hydrogen absorption features, spanning the entire bandpass.
\FUSE\ carriers no on-board source for wavelength calibrations and
consequently M27 has been observed numerous times for this purpose,
producing in the process a high quality data set as illustrated in
Figure~\ref{signos}.  The high density of the molecular lines presents
problems for the identification and analysis of atomic features
arising from the stellar photosphere and interstellar medium (ISM).
 
In a separate work \citep[][hereafter Paper I]{McCandliss:2007} we have
developed a molecular and atomic hydrogen absorption  model that, when
applied to a readily available pure hydrogen helium hot white dwarf
model of the CS \citep{Rauch:2003}, provides a reasonable template
against which the atomic absorption features from other metal line
systems can be isolated.  Here we compare the model to the
observations and present a list of line identifications for atomic
species associated with the stellar photosphere, the surrounding nebula
and non-nebular ISM.  We also present high resolution absorption line profiles
from STIS and \FUSE\ for the metals \ion{C}{1} \ion{-}{4}, \ion{N}{1}
\ion{-}{3}, \ion{O}{1}, \ion{O}{6}, \ion{Si}{2} \ion{-}{4}, \ion{P}{2}
\ion{-}{5}, \ion{S}{2} \ion{-}{4}, \ion{Ar}{1} \ion{-}{2} and
\ion{Fe}{2} \ion{-}{3}.

\begin{deluxetable*}{cccrrrr}
\tablecolumns{7}
\tablewidth{0pc}
\tablecaption{FUSE Observation Summary \label{fuseobslog} }
\tablehead{ \colhead{Observation} &\colhead{Date} & \colhead{APER} & \colhead{EXP:LiF1a}  & \colhead{EXP:SiC1b} & \colhead{EXP:LiF2a} & \colhead{EXP:SiC2a}\\
& & & \colhead{(ksec)}& \colhead{(ksec)} & \colhead{(ksec)}& \colhead{(ksec)} }
\startdata
M1070301&    2000-09-03&  MDRS&		15.5&	8.0&	15.6&	8.9\\
M1070302&    2000-09-24&  HIRS&		17.5&	4.8&	13.3&	5.6\\
M1070303&    2001-05-28&  LWRS&		7.4&	7.5&	7.3&	7.0\\
M1070304&    2001-05-29&  MDRS&		7.0&	4.8&	6.6&	0.6\\
M1070305&    2001-05-29&  HIRS&		8.0&	4.1&	6.3&	2.0\\
M1070306&    2001-07-28&  LWRS&		7.1&	7.1&	7.2&	7.1\\
M1070307&    2001-07-28&  MDRS&		6.7&	3.5&	6.5&	5.1\\
M1070308&    2001-07-29&  HIRS&		8.5&	3.4&	6.3&	3.3\\
M1070309&    2001-08-01&  LWRS&		7.5&	7.6&	7.7&	7.4\\
M1070310&    2001-08-01&  MDRS&		6.6&	4.1&	6.4&	4.8\\
M1070311&    2001-08-01&  HIRS&		7.8&	0.6&	5.6&	0.8\\
M1070312&    2002-10-30&  LWRS&		5.0&	5.1&	4.8&	5.0\\
M1070313&    2002-10-30&  MDRS&		7.0&	3.3&	6.4&	3.0\\
M1070314&    2002-11-02&  HIRS&		3.3&	0.9&	2.4&	0.7\\
M1070315&    2004-05-25&  LWRS&		5.3&	5.3&	4.3&	4.4\\
M1070316&    2004-05-25&  MDRS&		4.0&	2.9&	4.9&	4.4\\
M1070317&    2004-05-25&  HIRS&		6.5&	3.2&	5.0&	4.3\\
M1070319&    2002-11-03&  HIRS&		2.7&	1.1&	2.0&	0.9\\
P1043301&    2000-06-05&  LWRS&		16.9&	16.5&	16.3&	16.9\\
\enddata
\end{deluxetable*}

Paper I finds the observed atomic and molecular velocity stratification
in the nebular outflow of M27 is challenging to explain in the context
of the standard interacting winds model for PNe \citep{Kwok:1978}.  A
similar challenge has been made by \citet{Meaburn:2005} \citep[see
also][]{Meaburn:2005a,Meaburn:2005b} who used high resolution
position-and-velocity spectroscopy of the optical emission lines to
derive the ionization kinematics of several objects.  They find that
ballistic ejection could have been more important than interacting
winds in shaping the dynamics of PNe. In the interacting winds model a
high speed ($\sim$ 1000 \kms) highly ionized radiation driven wind from
the hot star shocks and ionizes a slow moving ($\sim$ 10 \kms), high
density mostly molecular asymptotic giant branch (AGB) wind.  In Paper
I no evidence was found for a high speed radiation driven wind.  The
upper limit to the terminal expansion velocity is $\approx$ 70 \kms.
The atomic line profiles show that high ionization material appears at
low velocities ($\lesssim$ 33 \kms) and low ionization material appears
at high velocities (33 -- 65 \kms).  At the transition velocity (33
\kms) between these two regimes, a predominately neutral atomic medium
appears along with the vibrationally excited \Htwo.  The molecular
abudance in the nebula is low ($N(\Htwo)$/$N($\ion{H}{1}) $\ll$ 1).

The Paper I study found that, despite the close proximity of a hot
central star, the molecular hydrogen ro-vibration levels of the ground
state were close to thermal ($\sim$ 2000 K).  They found no detectable
levels of far-UV fluorescence of molecular hydrogen, although
\lya\ pumped fluorescence has been detected by \citet{Lupu:2006}.
Surprisingly, they also found that the line-of-sight was devoid of
dust.  The temperature implied by the molecular hydrogen suggests that
the diffuse nebular medium is too hot for dust to form.  They argue
that the clumpy medium seen in the CO map of \citet{Bachiller:2000},
which cast the [\ion{O}{3}] shadows seen by \citet{Meaburn:1993} and
\citet{O'Dell:2002}, are the source of the excited molecular hydrogen
for the diffuse nebular medium and may harbor depleted material
(dust).  A depletion analysis of the Fe, Si, and S in the nebular
outflow could provide information as to whether these clumps are
composed of depleted material that is being released into the diffuse
medium along with the molecular hydrogen.

The digitial spectra used in this study, are available at the following
url,\\ {\small \verb+http://www.pha.jhu.edu/~stephan/H2ools/M27kruk04/+}. \\ This
dataset will 
be useful for photospheric analyses of the central star and for studies
seeking to understand the kinematics of the atomic and molecular
abundances in the nebular outflow.

\section{Datasets}

\subsection{\FUSE}
\label{fusespec}
\FUSE\ spectra of the CS (MAST object ID GCRV12336) were
acquired under the observatory wavelength calibration program (ID:
M107).  The CS was observed numerous times through all three
focal-plane apertures:  large (30\arcsec)$^2$,  medium (20\arcsec
$\times$ 4\arcsec) and small (20\arcsec $\times$ 1.25\arcsec); referred
to respectively  as LWRS, MDRS, and HIRS. A summary of the observations
is given in Table~\ref{fuseobslog}.   All spectra were acquired in
TTAG mode.  Each observation was performed as a series of exposures,
ranging in length from 0.5 ksec to 3 ksec.  Each exposure produces 8
individual spectra; one from each of two detector segments for each of
the 4 optical channels.  Effective exposure times are listed in the
table for the more interesting segment in each channel.  Exposure times
occasionally differ slightly from one segment to another because the
detector high voltage in a given segment can be reduced temporarily in response
to event bursts or brief current spikes.  Exposure times also differ
from one channel to another as a result of channel misalignment
effects.  For a description of the detectors, channel alignment issues,
systems nomenclature and other aspects of the \FUSE\ instrument, see
\citet{Moos:2000} and \citet{Sahnow:2000}.  

The exposures were processed with the following procedure.
The raw data were calibrated using Calfuse 3.0.7.  The resulting spectra for a given aperture (LWRS, MDRS
or HIRS) were screened to identify the peak detected stellar flux.
Exposures with less than 40\% of the peak flux (a small fraction) were
discarded. The remaining were normalized to the peak flux, yielding an
effective exposure time.  Channel/segments were coaligned with a
cross-correlation procedure on narrow ISM lines. Spectra were shifted
by whole pixel units to avoid smoothing and combined by effective
exposure time weighted average.  At this point there is one spectrum
for each spectrograph aperture, divided among the 8 channel/segment
combinations.  The MDRS spectra were put on an absolute wavelength
scale by adjusting each channel/segment spectrum so that the H$_2$
lines arising from the hot component of the molecular gas are at a
fiducial offset (--69 \kms, see \S~\ref{abslam} regarding this point).
Finally, the normalization of the MDRS and HIRS were readjusted to
match that of the LWRS and the zero point offsets of the HIRS and LWRS
channel/segment were adjusted to match the MDRS.\footnote{After
preforming these procedures spectral overlays the HIRS  LiF1b and LiF2a
segments were in obvious disagreement with the LWRS and MDRS  LiF1b and
LiF2a segments.  Consequently the HIRS wavelength scale was been
adjusted by adding in a linear dilation of the scale of 0.03 \AA\ for
every 45 \AA\ starting at 1125 \AA.  This adjustment to the HIRS wavelength scale has been incorporated into subsequent versions of CALFUSE.
The HIRS LiF2a flux was also
rescaled by a factor or 0.91 to achieve agreement with LiF1b.}  
We use the HIRS data in this work as it yields the highest resolution.

Each extracted segment has associated with it a set of one dimensional
arrays: wavelength (\AA), flux (\ergs), and estimated statistical error
(in flux units).  Figure~\ref{signos} shows the signal-to-noise and
spectral range for each of the coadded HIRS segments, obtained by
dividing a flux array by the corresponding statistical error array. The
signal-to-noise estimated in this way is purely statistical and does
not account for systemic errors, such as detector fixed pattern noise.
In the low sensitivity SiC channels the continuum signal-to-noise
ranges between  10 -- 25, while in the higher sensitivity LiF channels
it is  20 -- 55.  The resolution of the coadded spectra changes
slightly as a function of wavelength for each segment.  In modeling the
molecular hydrogen absorption, we find that a
gaussian convolution kernel with a full with at half maximum of 0.05
\AA\ at 1000 \AA\ (spectral resolution $R$ = 18,000, velocity
resolution ${\delta}V$ $\approx$ 17 \kms) provides a good match to the
unresolved absorption features throughout much of the bandpass.

Close comparison of the wavelength registration for overlapping segments
reveals isolated regions, a few \AA\ in length, of slight spectral
mismatch ($\sim$ a fraction of a resolution element) in the wavelength
solutions.  Consequently, combining all the spectra into one master
spectrum will result in a loss of resolution.  However, treating each
channel/segment individually increases the bookkeeping associated with
the data analysis.  Further, because fixed pattern noise tends to
dominate when the signal-to-noise is high, there is little additional
information to be gained in analyzing a low signal-to-noise data set
when high signal-to-noise is available.  For these reasons we elected
to form two spectra each of which covers the 900 -- 1190 bandpass
contiguously, using the following procedure.  The flux and error arrays
for the LiF1a and LiF1b segments were interpolated onto a common linear
wavelength scale with a 0.013 \AA\ bin, covering 900 -- 1190 \AA.  The
empty wavelength regions, being most of the short wavelength region
from 1000 \AA\ down to the 900 \AA\ and the short gap region in between
LiF1a and LiF1b, were filled in with most of SiC2a and a small portion
of SiC2b respectively.  We refer to this spectrum as s12.  The LiF2b,
LiF2a, SiC1a and SiC1b segments were merged similarily into a spectrum,
s21.  Absorption line analyses were carried out using both s12 and s21.

\subsection{STIS}
\label{stis}
The E140M spectrum of M27 (o64d07020\verb'_'x1d.fits), taken from the
Multimission Archive at Space Telescope (MAST),  was acquired for
\HST\ Proposal 8638 (Klaus Werner -- PI).  These data are a high
level data product consisting of arrays of flux and wavelength
calibrated, one-dimensional extractions of individual echelle orders.
The data were acquired through the 0\farcs2$\times$0\farcs2
spectroscopic aperture for an exposure time of 2906 s and were reduced
with CALSTIS version 2.18.  The spectral resolution of the E140M is
given in the STIS data handbook \citep[version 7.0][]{Quijano:2003} as
$R$ $\approx$ 48,000 (${\delta}V$~=~6.25~\kms).  The intrinsic line
profile for the 0\farcs2$\times$0\farcs2  aperture has a gaussian core
with this width, but there is non-negligible power in the wings.
These data offer a high velocity resolution view of the ionization stratification in the nebular expansion.

\begin{figure*}
\figurenum{2a}
\includegraphics[height=9.in]{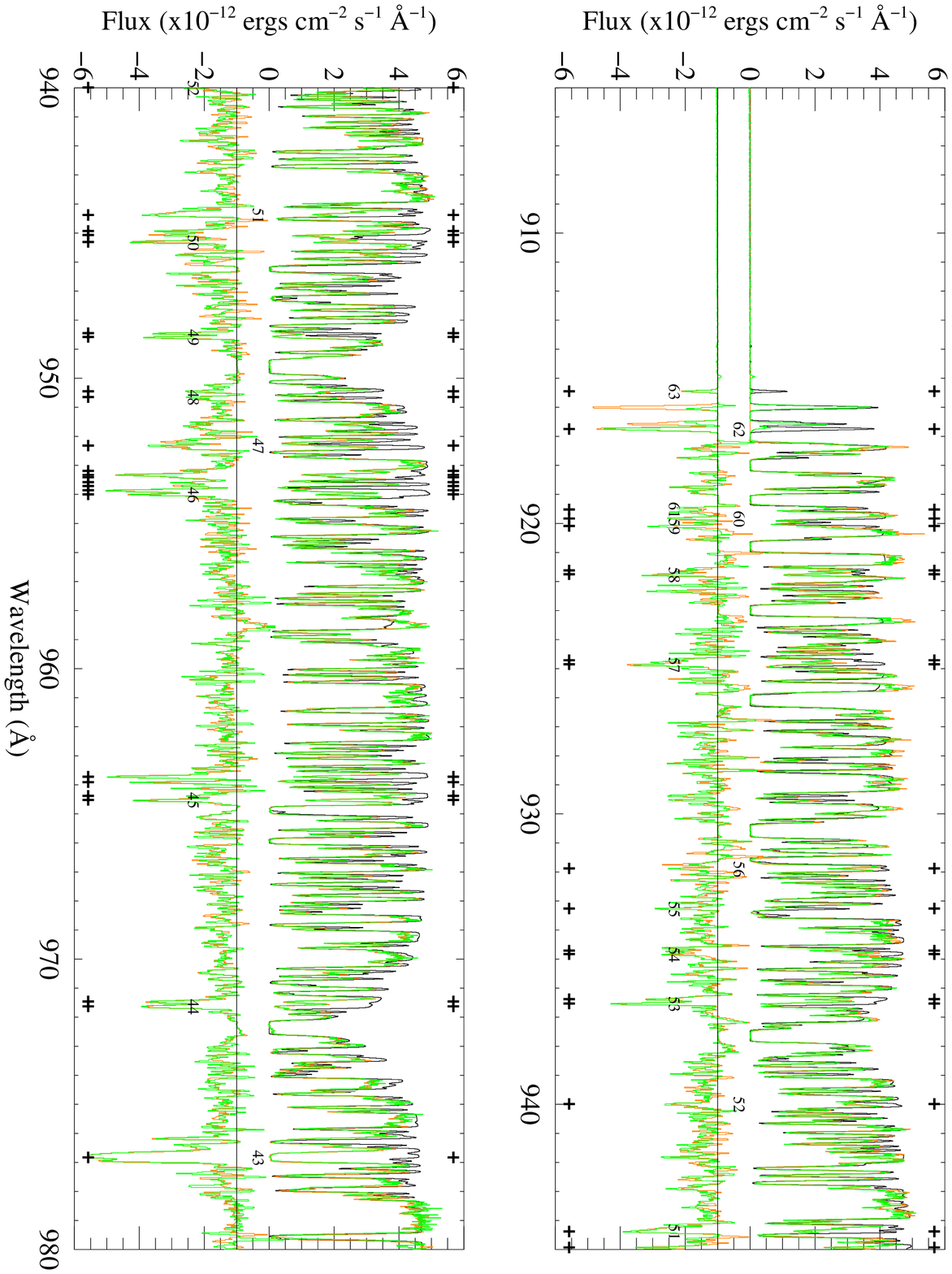}
\vspace*{-.5in}
\figcaption[f2a.eps, 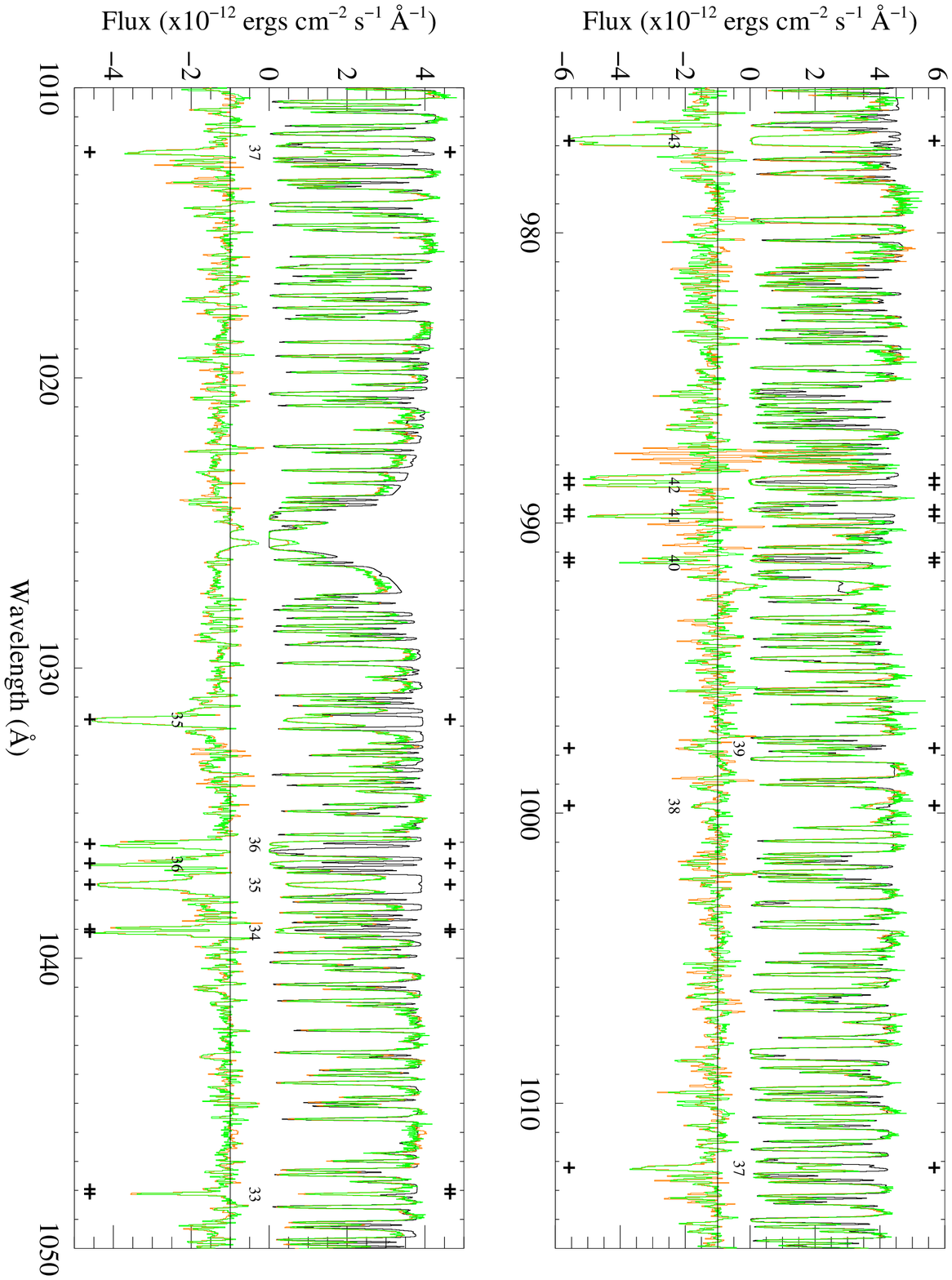, 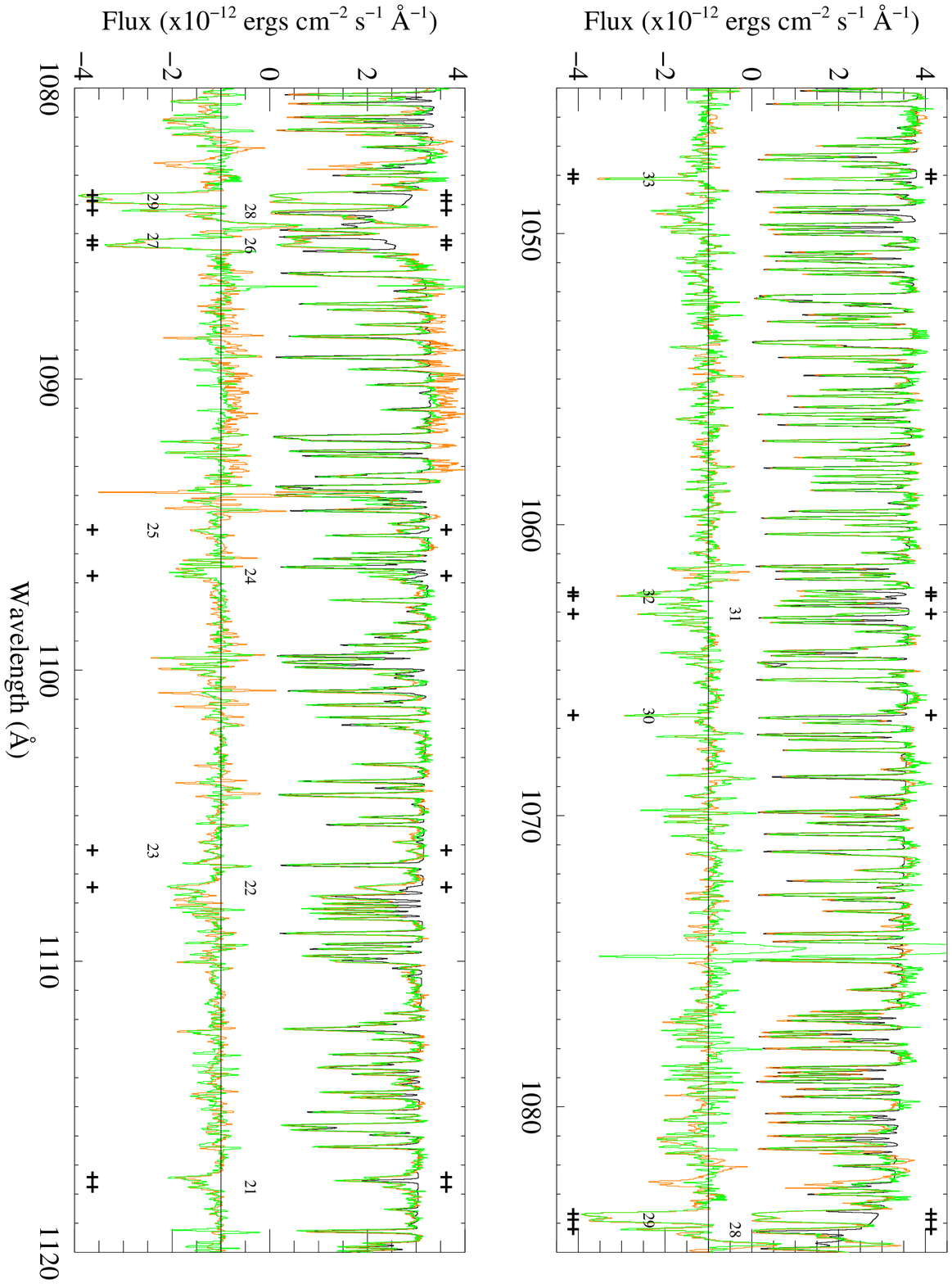, 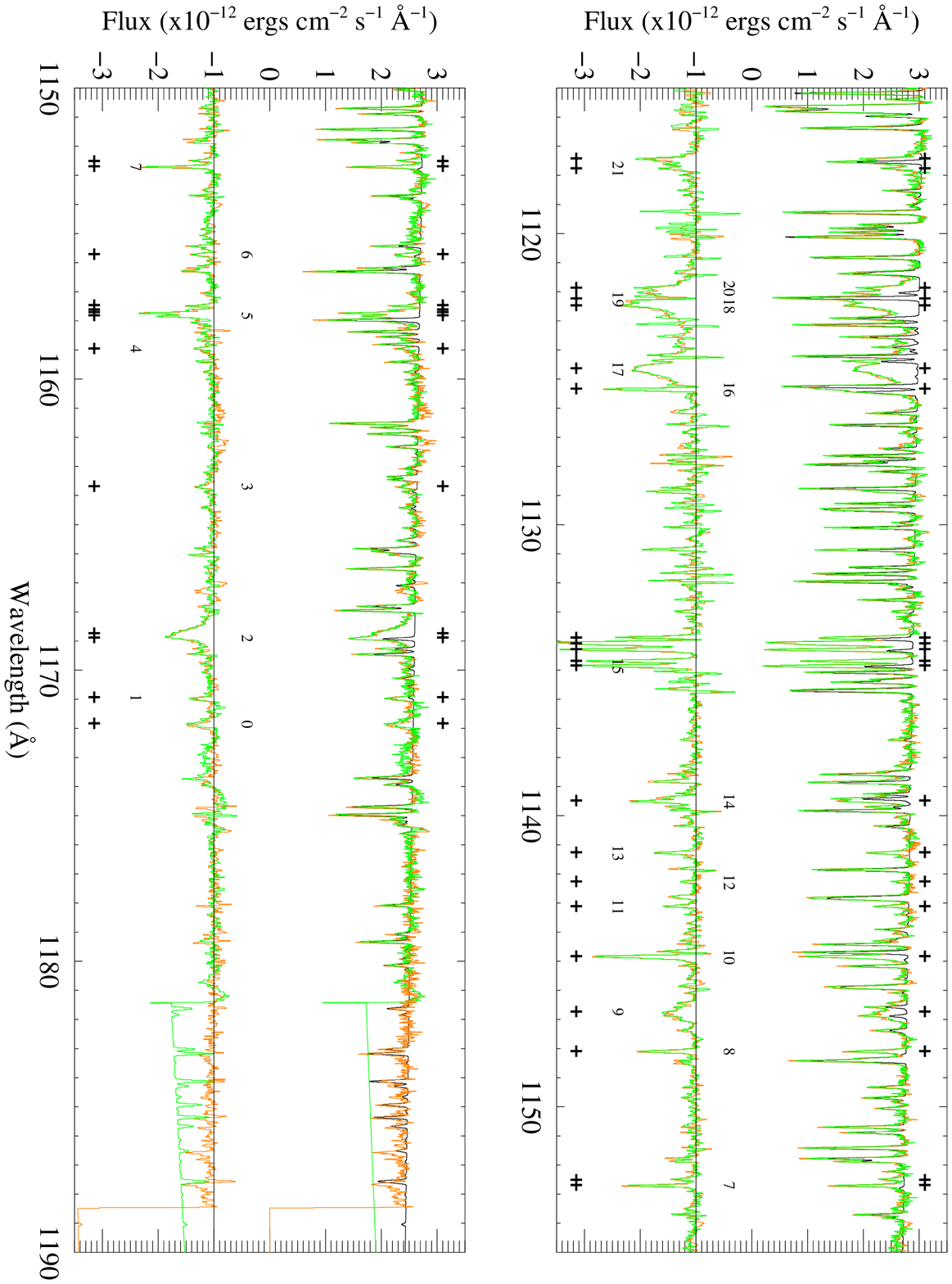]{ \label{residual0}
Hydrogen absorbed continuum model 905 -- 980 \AA. Identified line blends are numbered and marked
with upper and lower + and tabulated in Table ~\ref{lineid}. Upper
portion of each panel is an overplot of the s12 (orange) and s21
(green) \FUSE\ spectra,  and the model (black). Lower portion of each
panel shows the data after subtraction of the model, offset by
10$^{-12}$ \ergs.
}
\end{figure*}

\begin{figure*}
\figurenum{2b}
\includegraphics[height=9in]{f2b.eps}
\vspace*{-.5in}
\figcaption[f2b.eps]{Hydrogen absorbed continuum model 975 -- 1050 \AA.  See Figure 2a for a description of the colors and symbols.}
\end{figure*}

\begin{figure*}
\figurenum{2c}
\includegraphics[height=9in]{f2c.eps}
\vspace*{-.5in}
\figcaption[f2c.eps]{Hydrogen absorbed continuum model 1045 -- 1120 \AA.
See Figure 2a for a description of the colors and symbols.}
\end{figure*}

\begin{figure*}
\figurenum{2d}
\includegraphics[height=9in]{f2d.eps}
\vspace*{-.5in}
\figcaption[f2d.eps]{Hydrogen absorbed continuum model 1115 -- 1190 \AA.
See Figure 2a for a description of the colors and symbols.}
\end{figure*}

\setcounter{figure}{2}

\section{Model of Continuum Absorption by Hydrogen and Line Identifications}
\label{contmod}

A model of the CS  SED, including the atomic and molecular hydrogen
absorptions has been developed by \citet{McCandliss:2007} to aid
identification of photospheric, nebular or non-nebular velocity
components in the \FUSE\ and STIS metal line profiles.  We used
synthetic stellar flux interpolated from the grid of \citet{Rauch:2003}
with $\log{g}$ = 6.5, T = 120,000 K, and ratio of H/He = 10/3 by mass.
This model includes no metals and is consistent with, although slightly
hotter than, the quantitative spectroscopy of \citet{Napiwotzki:1999}.
The adopted temperature and gravity has slightly less pressure
broadening and is a better match to the observed Lyman lines towards
the series limit.  We also adopt a stellar mass of 0.56 $M_{\sun}$ as
suggested by post asymptotic giant branch evolutionary tracks.  Use of
this mass along with the above gravity required a distance of 466 pc to
match the absolute flux, which is at the upper limit given by
\citet{Benedict:2003}.  Beyond our immediate need to match the absolute
flux for the given gravity and mass there is no particular reason to
prefer this distance over that derived by Benedict et al.  Questions
regarding the acceptable uncertainty in distance, absolute flux and
derived stellar parameters are best left for a stellar model
specifically tailored to include the effects of metals, gravity and
evolutionary state.  The data supplied here is intended to enable such an effort.

\LongTables

\begin{deluxetable}{rrcc}
\tablecolumns{4}
\tablewidth{0pc}
\tablecaption{Stellar, Nebular and Non-Nebular Line ID's \label{lineid}}
\tablehead{ \colhead{Feature} &\colhead{$\lambda_{obs}$} & \colhead{Atomic 
ID, $\lambda_{rest}$ } & \colhead{Component\tablenotemark{$\dagger$}}  \\
	& \colhead{\AA}	& \colhead{\AA}	&	}
\startdata
0	&1171.85&\ion{O}{6} 1172.439?		&Ph?	\\
1	&1170.95&\ion{O}{6} 1171.561?		&Ph?	\\
2	&1168.90&\ion{C}{4} 1168.993		&Ph	\\
\ldots	&1168.75&\ion{C}{4} 1168.849		&Ph	\\
3	&1163.70&?				&?	\\
4	&1158.97&\ion{P}{2}** 1159.087	 	&NN	\\
5	&1157.83&\ion{C}{1} UV15.01+16		&Nb,NN\\
\ldots	&1157.72&\ion{C}{1} UV15.01+16		&Nb,NN\\
\ldots	&1157.63&\ion{C}{1} UV15.01+16		&Nb,NN\\
\ldots	&1157.48&\ion{C}{1} UV15.01+16		&Nb,NN\\
6	&1155.73&\ion{C}{1} UV19		&Nb,NN\\
7	&1152.71&\ion{P}{2} 1152.818		&NN\\
\ldots	&1152.53&\ion{P}{2} 1152.818		&Nb\\
8	&1148.11&?				&?	\\
9	&1146.75&\ion{O}{6} 1147.072 + 1146.791	&Ph\\
10	&1144.85&\ion{Fe}{2} 1144.938		&NN\\
11	&1143.13&\ion{Fe}{2} 1143.226			&NN\\
12	&1142.28&\ion{Fe}{2} 1142.366			&NN\\
13	&1141.28&?					&?	\\
14	&1139.50&\ion{C}{1} 1139.793 (UV22)		&Nb\\
15	&1134.86&\ion{N}{1} 1134.980			&NN\\
\ldots	&1134.70&\ion{N}{1} 1134.980			&Nb\\
\ldots	&1134.30 &\ion{N}{1} 1134.415			&NN\\
\ldots	&1134.10 &\ion{N}{1} 1134.415 + 1134.1653&Nb+NN\\
\ldots	&1133.90 &\ion{N}{1} 1134.165			&Nb\\
16	&1125.35&\ion{Fe}{2} 1125.448			&NN\\
17	&1124.65&\ion{O}{6} 1124.809 +1124.716		&Ph\\
18	&1122.49&\ion{O}{6} 1122.618 + \ion{C}{1} UV27&Ph,Nb\\
19	&1122.25&\ion{O}{6} 1122.348		&Ph,Nb\\
20	&1121.88&\ion{Fe}{2} 1121.975			&NN\\
21	&1117.78&\ion{C}{1} UV29		&Nb,NN\\
\ldots	&1117.44&\ion{C}{1} UV29		&Nb,NN\\
22	&1107.48&\ion{C}{4} 1107.593		&Ph	\\
23	&1106.21&\ion{C}{1} 1106.316		&NN\\
24	&1096.78&\ion{Fe}{2} 1096.877			&NN\\
25	&1095.20&?				&?	\\
26	&1085.42&\ion{N}{2}** 1085.710		&Nb\\
27	&1085.25&\ion{N}{2}** 1085.551 + 1085.533	&Nb\\
28	&1084.23&\ion{N}{2}* 1084.584 + 1084.566&Nb\\
29	&1083.90&\ion{N}{2} 1083.994		&NN\\
\ldots	&1083.70&\ion{N}{2} 1083.994		&Nb\\
30	&1066.57&\ion{Ar}{1} 1066.660		&NN\\
31	&1063.09&\ion{Fe}{2} 1063.176		&NN\\
32	&1062.47&\ion{S}{4}  1062.664		&Ph,Nb\\
\ldots	&1062.35&\ion{S}{4}  1062.664		&Nb\\
33	&1048.14&\ion{Ar}{1} 1048.220		&NN\\
\ldots	&1047.98&\ion{Ar}{1} 1048.220		&Nb\\
34	&1039.12&\ion{O}{1} 1039.230		&NN\\
\ldots	&1039.02&\ion{O}{1} 1039.230		&Nb\\
35	&1037.48&\ion{O}{6} 1037.617		&Ph\\
36	&1036.75&\ion{C}{2}* 1037.018		&Nb\\
36	&1036.08&\ion{C}{2} 1036.337		&Nb\\
35	&1031.79&\ion{O}{6} 1031.926		&Ph\\
37	&1012.24&\ion{S}{3} 1012.495		&Nb\\
38	& 999.76&?				&?	\\
39	& 997.78&\ion{P}{3} 998.000		&Nb\\
40	& 991.38&\ion{N}{3} 991.577 + 991.511	&Nb\\
\ldots	& 991.22&\ion{N}{3} 991.577 + 991.511	&Nb?\\
41	& 989.79&\ion{Si}{2} 989.873		&NN\\
\ldots	& 989.56&\ion{N}{3} 989.799		&Nb\\
42	& 988.71&\ion{O}{1} 988.773		&NN\\
\ldots	& 988.47&\ion{O}{1} 988.773 + 988.6549	&Nb,NN\\
43	& 976.85&\ion{C}{3} 977.020	&Ph,Nb,NN\\
44	& 971.65&\ion{O}{1} 977.738		&NN\\
\ldots	& 971.48&\ion{O}{1} 977.738		&Nb\\
45	& 964.54&\ion{N}{1} 964.626		&NN\\
\ldots	& 964.40&\ion{N}{1} 964.626		&Nb\\
\ldots	& 963.93&\ion{N}{1} 963.990		&NN\\
\ldots	& 963.74&\ion{N}{1} 963.990 +\ion{P}{2} 963.801	&Nb\\
46	& 954.02&\ion{N}{1} 954.104		&NN\\
\ldots	& 953.88&\ion{N}{1} 954.104 + 953.970	&Nb,NN\\
\ldots	& 953.73&\ion{N}{1} 953.970		&Nb\\
\ldots	& 953.60&\ion{N}{1} 953.655		&NN\\
\ldots	& 953.43&\ion{N}{1} 953.655		&Nb\\
\ldots	& 953.35&\ion{N}{1} 953.415		&NN\\
\ldots	& 953.21&\ion{N}{1} 953.415		&Nb\\
47	& 952.34&\ion{N}{1} 952.415		&NN\\
48	& 950.67&\ion{O}{1} 950.885		&Nb\\
\ldots	& 950.47&\ion{P}{4} 950.657		&Nb+(Ph?)\\
49	& 948.60&\ion{O}{1} 948.686		&NN\\
\ldots	& 948.47&\ion{O}{1} 948.686		&Nb\\
50	& 945.33&\ion{C}{1} 945.579		&Nb\\
\ldots	& 945.09&\ion{C}{1} 945.338		&Nb\\
\ldots	& 944.95&\ion{C}{1} 945.191		&Nb\\
51	& 944.40&\ion{S}{6} 944.523		&Nb+Ph\\
52	& 940.01&?				&?	\\
53	& 936.55&\ion{O}{1} 936.623		&NN\\
\ldots	& 936.40&\ion{O}{1} 936.623		&Nb\\
54	& 934.85&?				&?	\\
\ldots	& 934.73&?				&?	\\
55	& 933.28&\ion{S}{6} 933.378?		&?	\\	
56	& 931.90&\ion{Ar}{2}* 932.054		&Nb\\
57	& 924.87&\ion{O}{1} 924.950		&NN\\
\ldots	& 924.73&\ion{O}{1} 924.950		&Nb\\
58	& 921.77&\ion{O}{1} 921.857		&NN\\
\ldots	& 921.64&\ion{O}{1} 921.857		&Nb\\
59	& 920.11&?				&?	\\
60	& 919.85&?				&?	\\
61	& 919.54&\ion{Ar}{2} 919.781?,\ion{O}{1} 919.658?	&?\\
62	& 916.77&\ion{O}{1} 916.960+916.815	&Nb,NN\\
63	& 915.47&\ion{N}{2} 915.61		&Nb,NN\\
\enddata
\tablenotetext{$\dagger$}{Ph = Photosphere, Nb = Nebular, NN = 
Non-Nebular, ? = Unknown  }
\end{deluxetable}

The model atomic and molecular absorption spectrum was created by adding
to a master optical depth template, covering 900 -- 1400 \AA, all  the
contributions from the atomic and molecular hydrogen lines for all the
velocity components.  There are two distinct components with
heliocentric velocities of --75  and --28   $\pm$ 2 \kms\ for molecular
hydrogen. The first is associated with the nebular expansion and the
second is associated with non-nebular ISM.  For the atomic Lyman series
we used transitions up to $n$ = 49, with the  velocity offsets, column
densities and doppler parameters listed in \citet{McCandliss:2007}.  For  hot molecular hydrogen we used the
H$_2$ools rovibrational templates of \citet{McCandliss:2003} with b = 7
\kms, offset by --75 \kms and scaled to the column densities listed in
\citet{McCandliss:2007}. We extended the calculation
up to \Jpp = 15 and \vpp = 3  by assuming the upper level ro-vibration
populations were thermal with a temperature of 2040 K as given by the best fit single temperature model.  We also used the  H$_2$ools templates to
model the cold molecular component at --28 \kms\ with a pure thermal
distribution of 200 K.  The transmission function, defined as the
negative exponential of the master optical depth template, is
multiplied to the stellar continuum model, which has been blueshifted
to the systemic radial velocity of the nebula, --42 \kms, to produce the
model of continuum absorption by hydrogen.  This model is also made available
on the H$_2$ools website.

\begin{figure*}
\includegraphics[height=8.5in]{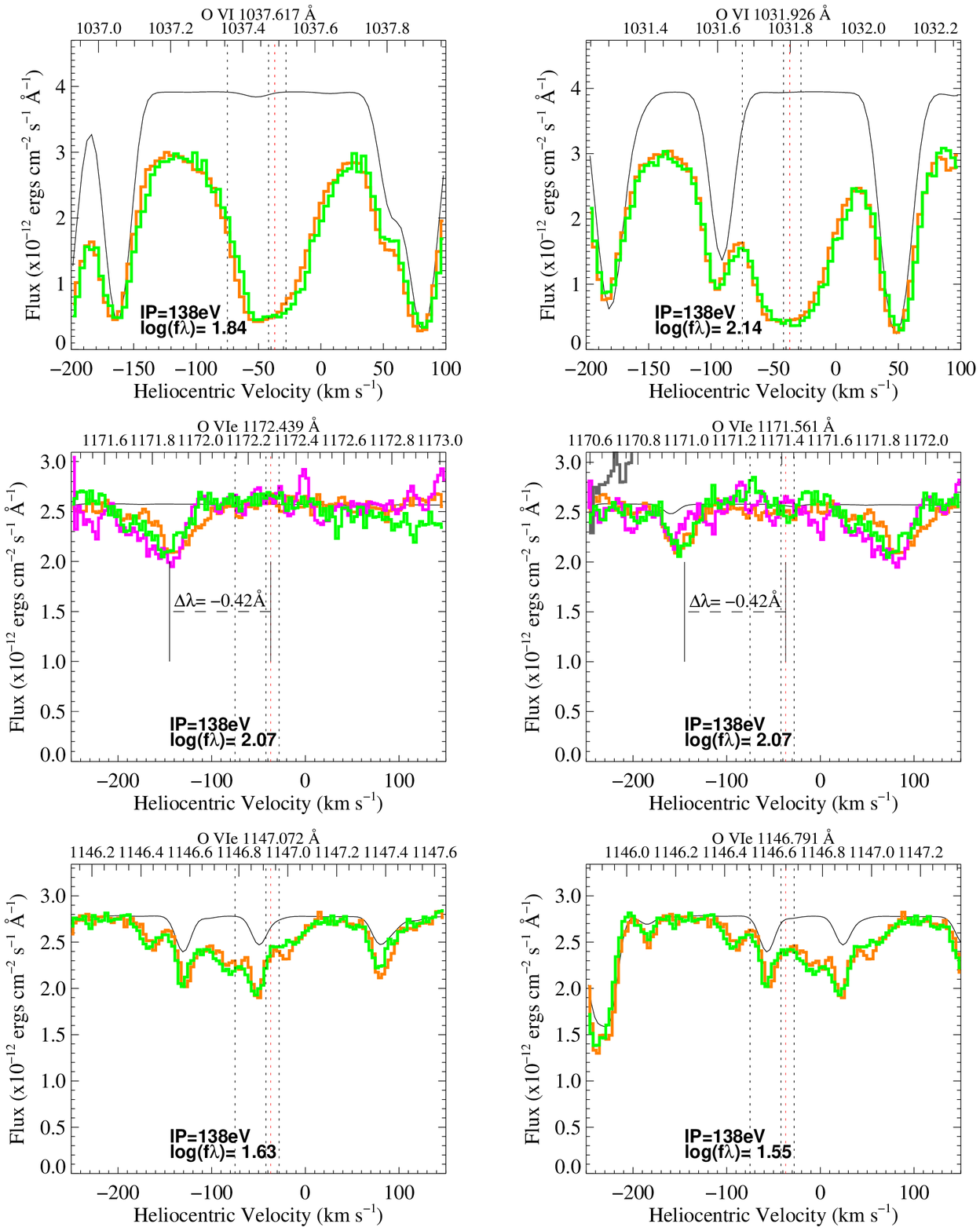}
\figcaption[f3.eps]{\label{oviov0}
Stellar photospheric \ion{O}{6} and \ion{O}{6}e lines. \FUSE\ spectra
s12 (orange), s21 (green) and the model (black) are plotted along with
overlapping STIS orders (purple and grey), if available.  Vertical
dashed lines mark the transition, $V_{sys}$, $V_{gr}$ and non-nebular
velocities at --75, --42, --37 and --28 \kms\ respectively.
\ion{O}{6}e \dlam 1172.439, 1171.561 lines were used to assess the
offset between the \FUSE\ and STIS spectra.  Published rest wavelengths
for these lines are discrepant.
 }
\end{figure*}

\begin{figure*}
\includegraphics[height=8.5in]{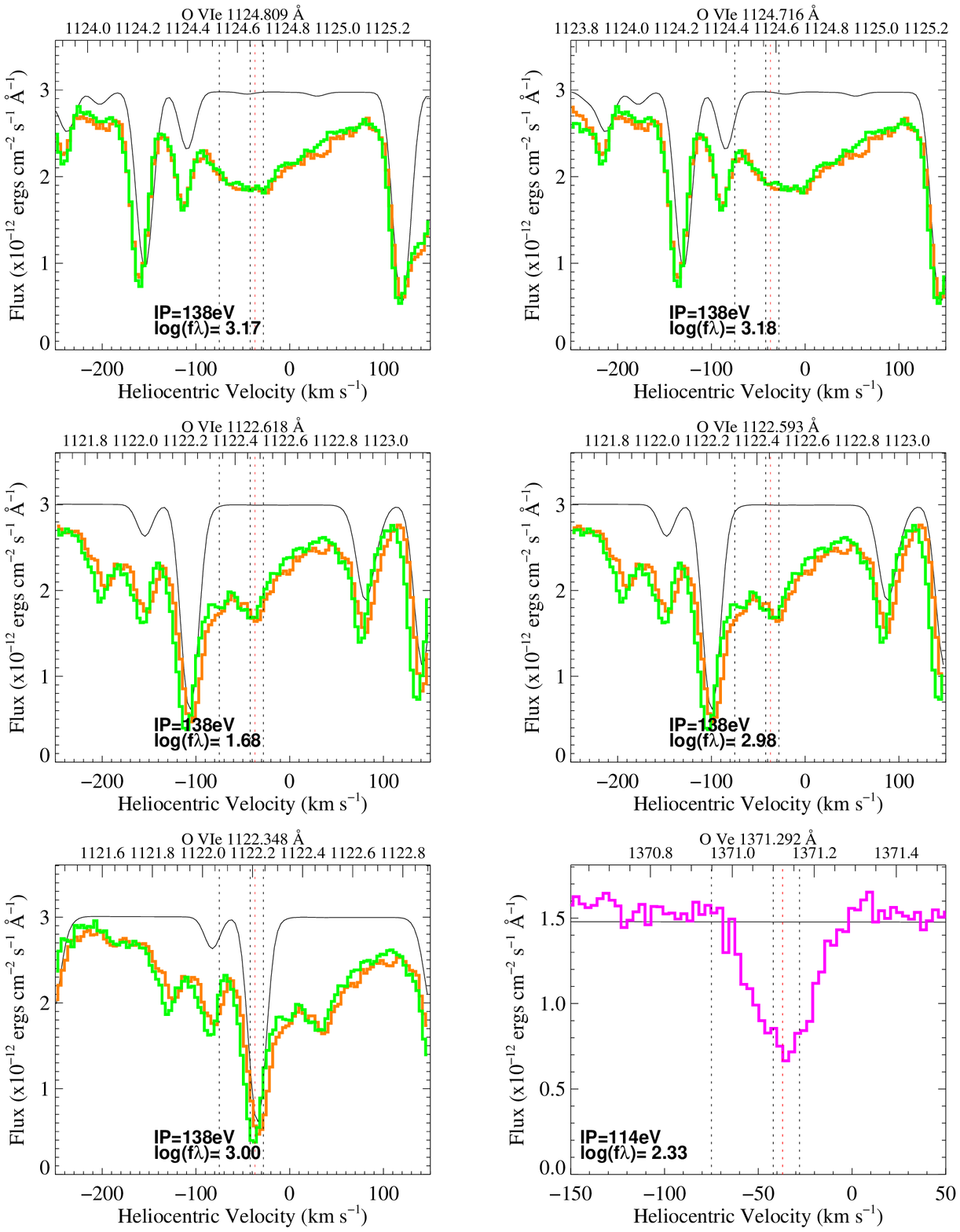}
\figcaption[f4.eps]{ \label{oviov1}
Stellar photospheric \ion{O}{6}e and \ion{O}{5}e \lam1371.292 lines.
Stellar photospheric \ion{O}{6}e and \ion{O}{5}e \lam1371.292 lines.
The \ion{O}{5}e \lam1371.292 line was used to established the
gravitational redshift of the system.  See Figure~\ref{oviov0} for a
description of the colors.
}
\end{figure*}

\begin{figure*}
\includegraphics[height=8.5in]{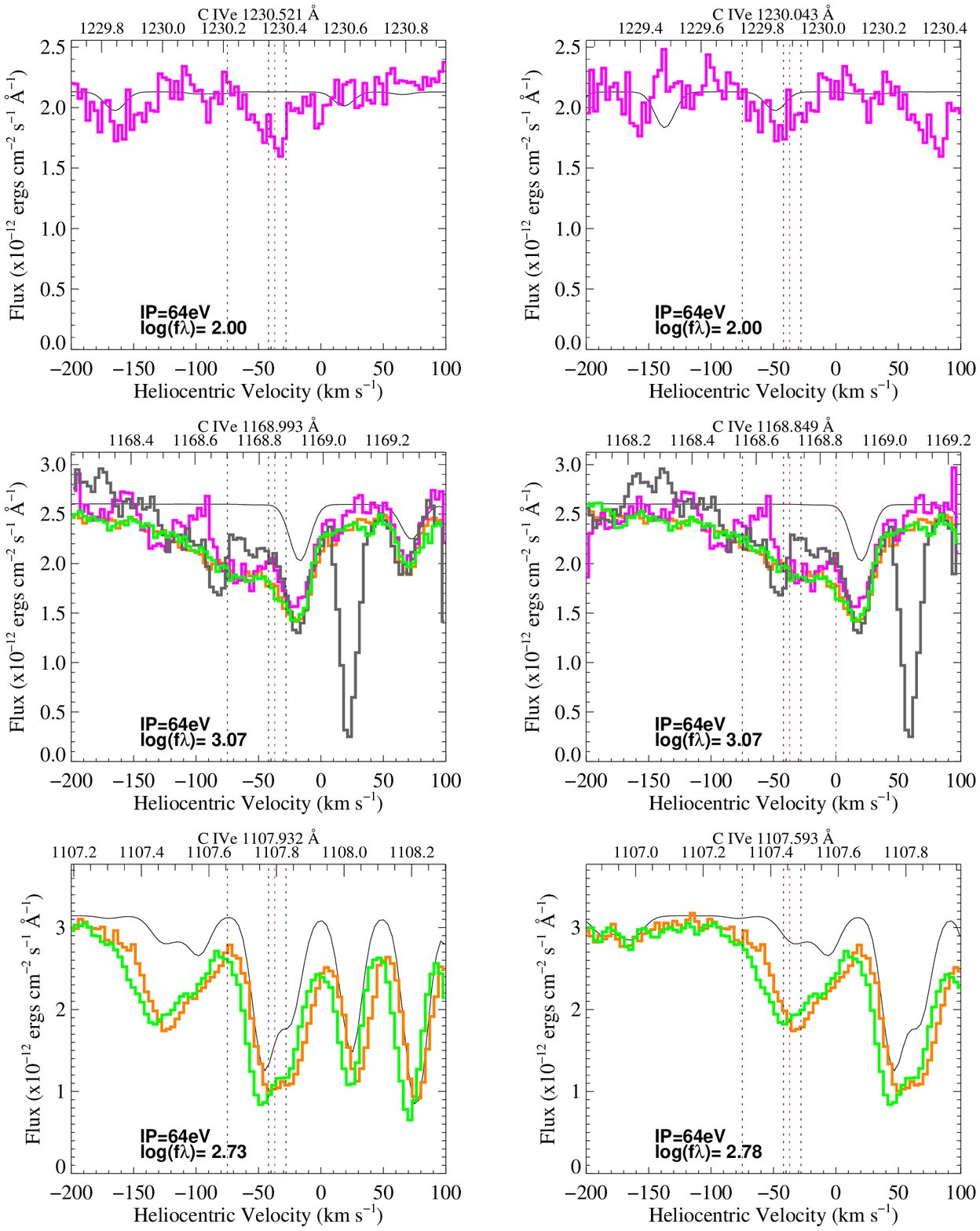}
\figcaption[f5.eps]{ \label{cive}
The photospheric lines of excited \ion{C}{4}e.  \ion{C}{4}e \dlam
1168.849, 1168.993 were used to reconcile the relative offset between
the \FUSE\ and STIS wavelength scales.  See Figure~\ref{oviov0} for a
description of the colors.
}
\end{figure*}

Figure~\ref{residual0} shows the resulting continuum absorption model.
The top spectral sequence in each panel is an overplot of the s12
(orange) and s21 (green) spectral extractions with the absorbed
continuum model (black).  The lower sequence in each panel is the
residual formed by substracting the model from s12 (orange) and s21
(green).  An offset of 10$^{-12}$ \ergs\ and a two pixel smoothing has
been applied to the residuals.   The solid horizonal black line marks
the zero level for the residuals.  The high frequency spikes that go
rapidly from positive to negative in the residual are created by local
wavelength misalignment of the spectral extractions with respect to the
molecular model.  The strongest stellar  photospheric, nebular and
non-nebular absorptions by species other than \Htwo, \ion{H}{1} or
photospheric \ion{He}{2} are clearly revealed in the residuals.
Emission lines of \ion{He}{2} \lam 1084, \lam 992 and \lam 958 are also
evident.  The absorption systems are numbered in the figure in
reference to an entry in Table~\ref{lineid} where the line
specifications are detailed.  The identifications are not exhaustive;
small unidentified lines may yet lurk in the noise.

The residual spectrum allows us to assess the success of the absorbed continuum
model in reproducing the features in the spectral
extractions.  Strictly speaking, the subtraction of features should
take place in the optical depth space $\tau(\lambda)$ as opposed to the
transmission space $\exp(-\tau(\lambda))$.  Abundance determinations
whether by profile fitting or equivalent width determination must
account for the optical depth blending correctly.  However, as our
purpose here is merely line identification, the simple subtraction
process is adequate.
 
\begin{figure*}[t] 
\includegraphics[height=8.5in]{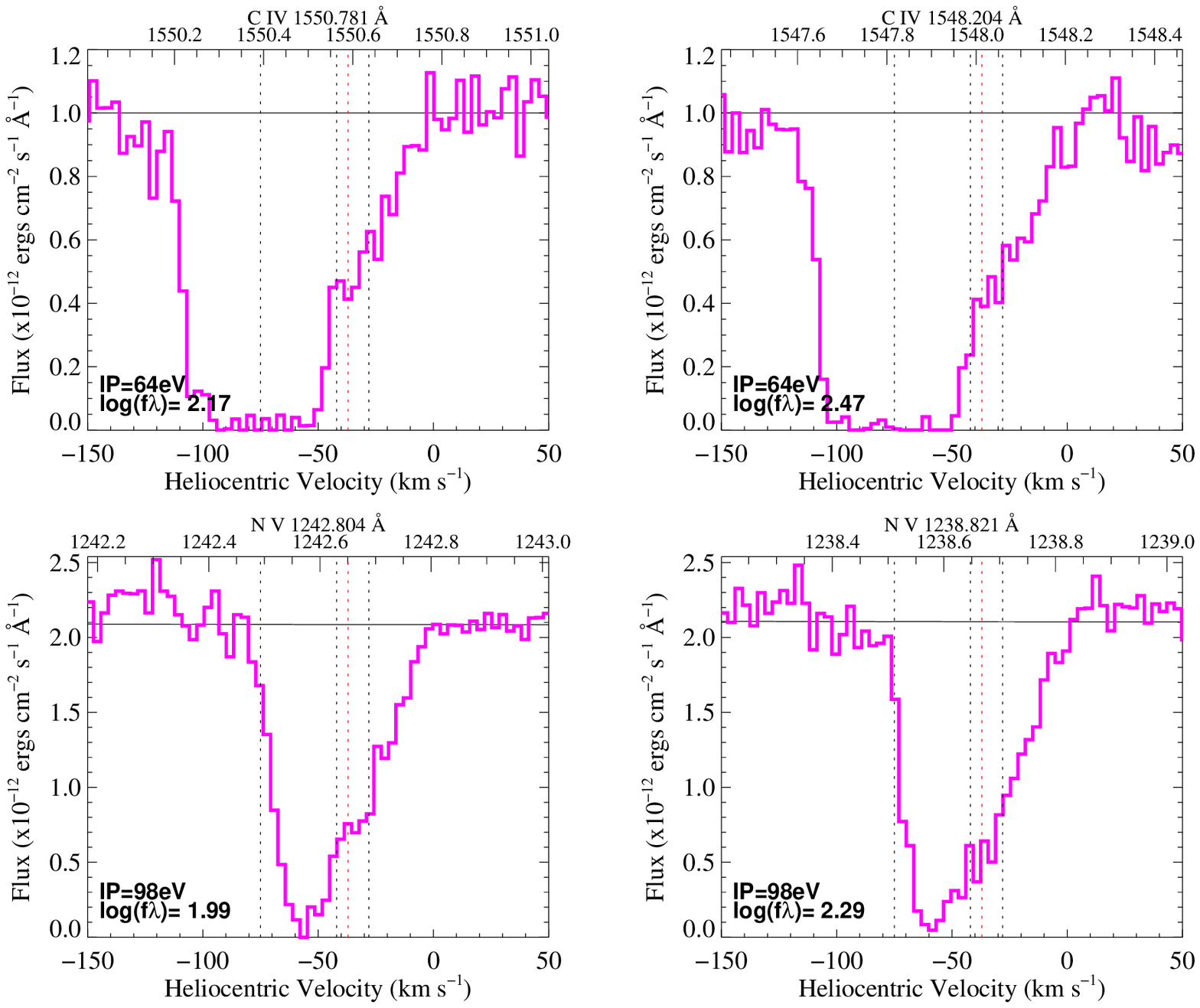}
\vspace*{-3in}
\figcaption[f6.eps]{ \label{civnv}
Absorption as a function of velocity for \ion{C}{4} \dlam 1548.204,
1550.781, and \ion{N}{5} \dlam 1238.821, 1242.804, showing signs of
nebular absorption blueward of --37 \kms\ and photospheric absorption
redward. See Figure~\ref{oviov0} for a description of the colors.
}
\end{figure*}

\begin{figure*}
\includegraphics[height=8.5in]{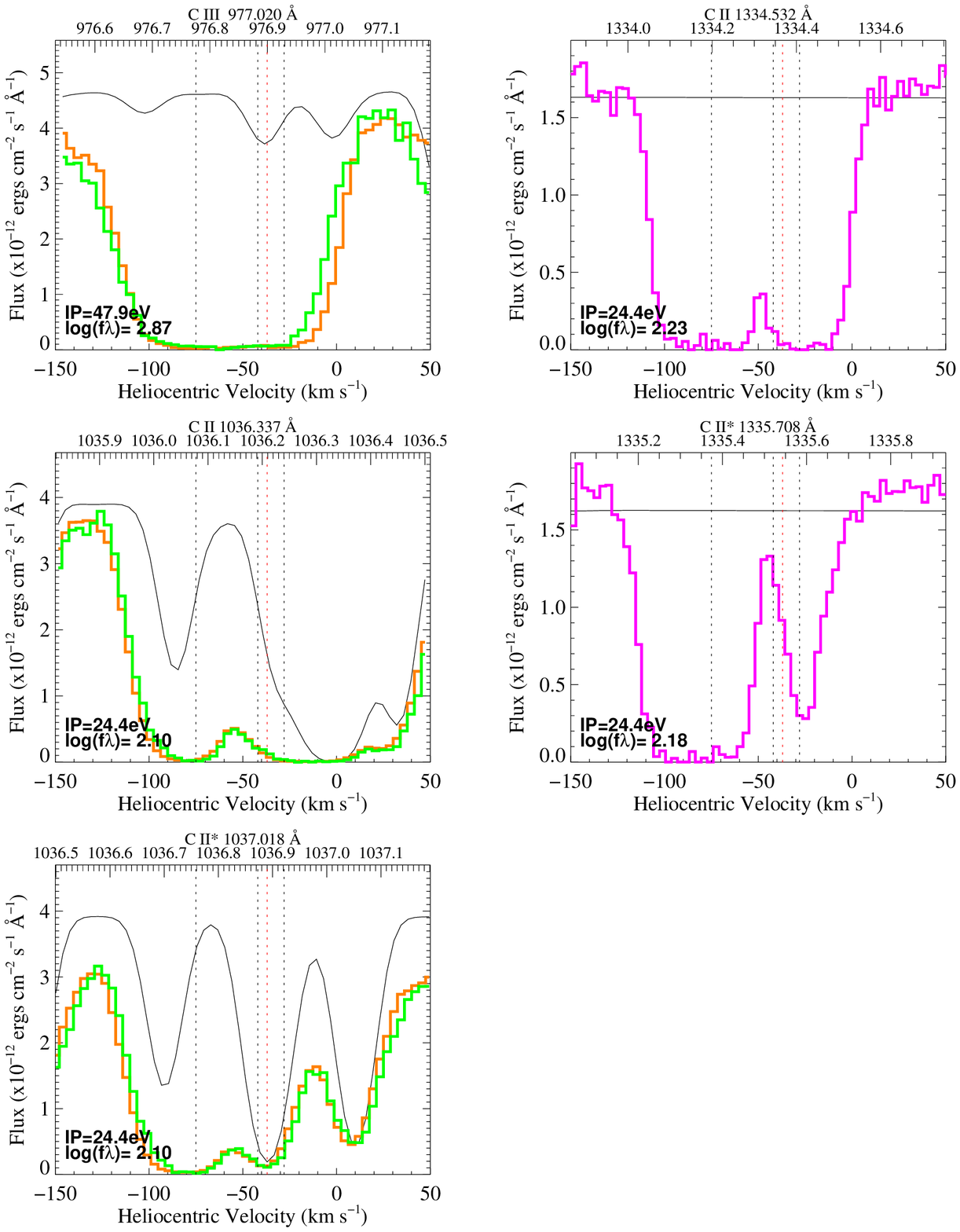}
\figcaption[f7.eps]{\label{ciiicii}
Absorption as a function of velocity for the \ion{C}{3} and \ion{C}{2}
lines.  Like the \ion{C}{4} lines, these lines are heavily saturated
throughout the nebular flow region blueward of  --42 \kms. See
Figure~\ref{oviov0} for a description of the colors.
} 
\end{figure*}

\begin{figure} \hspace*{-.35in}
\includegraphics[height=8.5in]{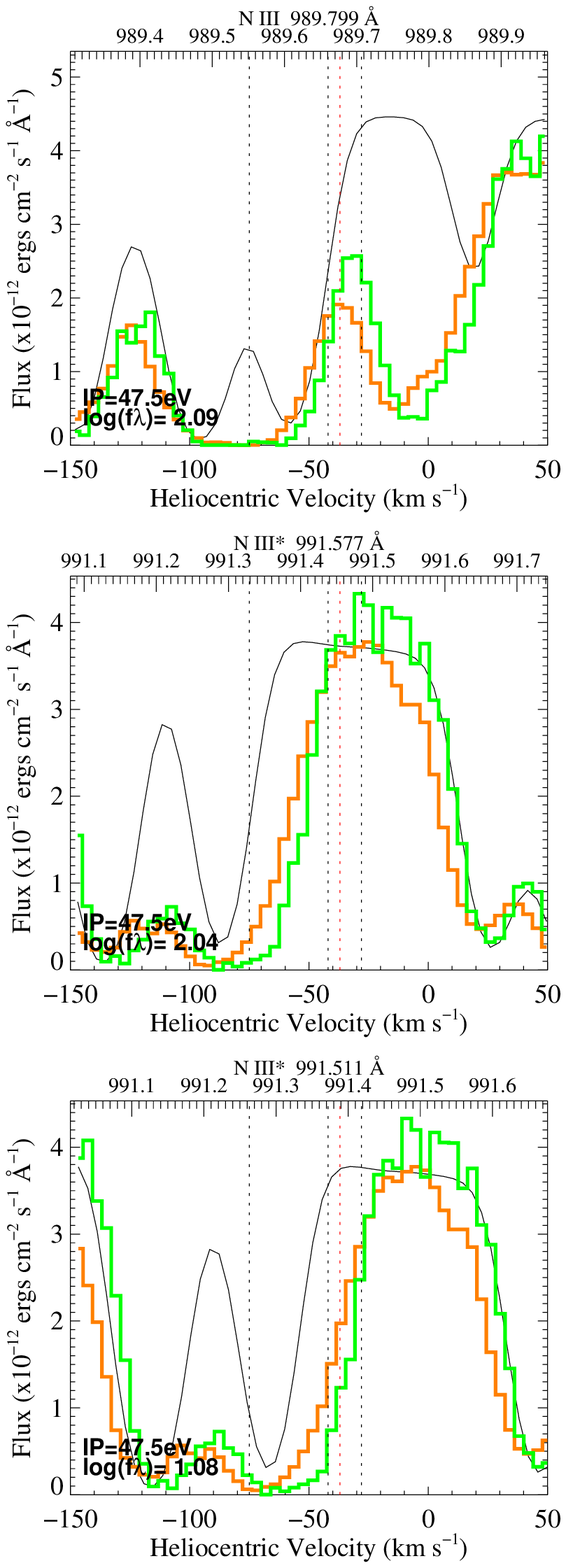}
\figcaption[f8.eps]{\label{niii}
Absorption as a function of velocity for the \ion{N}{3} \dlam 989.799
-- 991.577 multiplet.  The \ion{N}{3} is blended with molecular
hydrogen. The \ion{N}{3}* lines are also blended with each other.   See
Figure~\ref{oviov0} for a description of the colors.
} 
\end{figure}

\begin{figure*}
\includegraphics[height=8.5in]{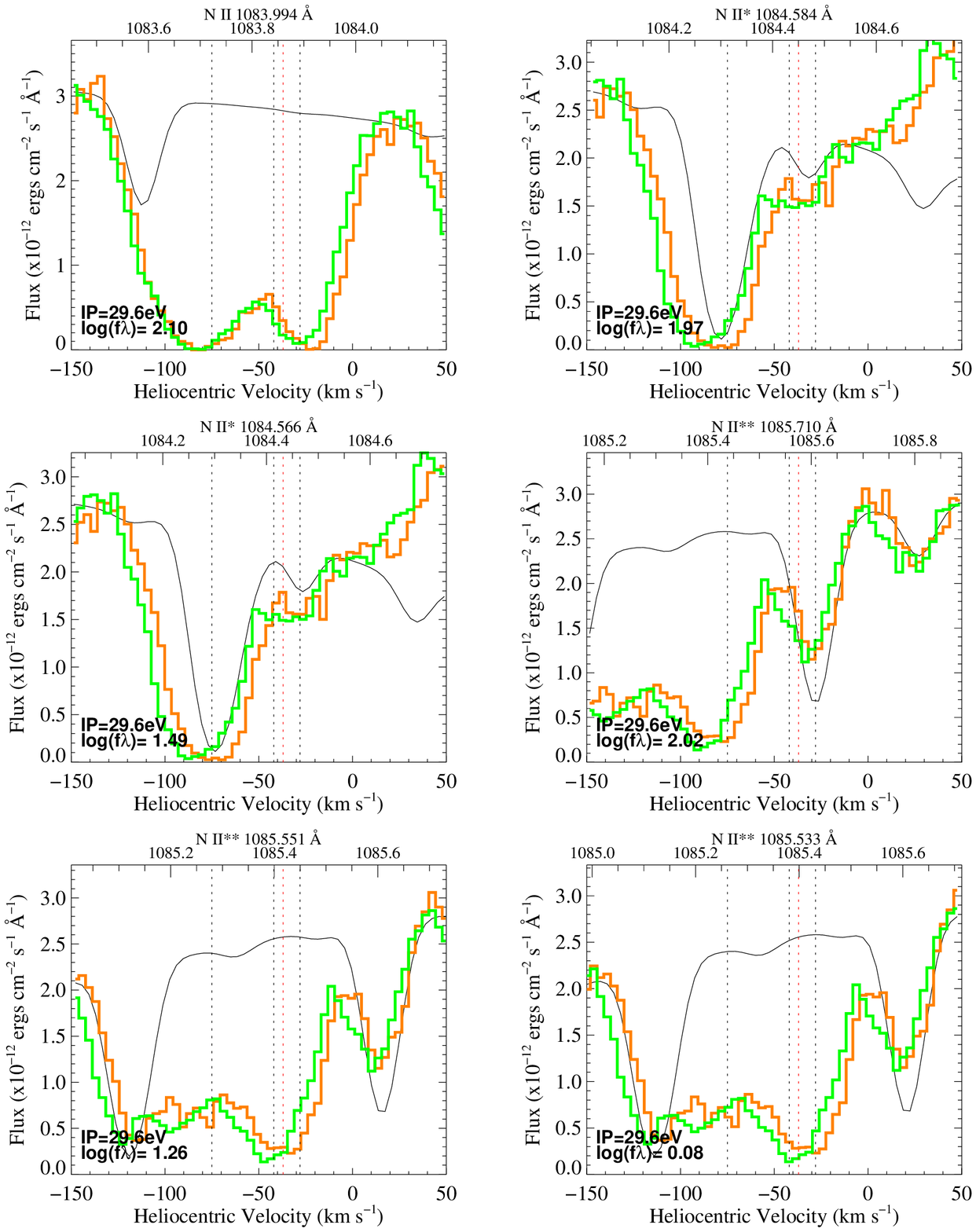}
\figcaption[f9.eps]{\label{nii}
Absorption as a function of velocity for the \ion{N}{2} \dlam 1083.994
-- 1085.710 multiplet.  The \ion{N}{2}* and \ion{N}{2}** lines are
blended with each other and molecular hydrogen.  The \ion{N}{2} line is
relatively clean.  See Figure~\ref{oviov0} for a description of the
colors.
} 
\end{figure*}

\begin{figure*}
\figurenum{10a}
\includegraphics[height=8.5in]{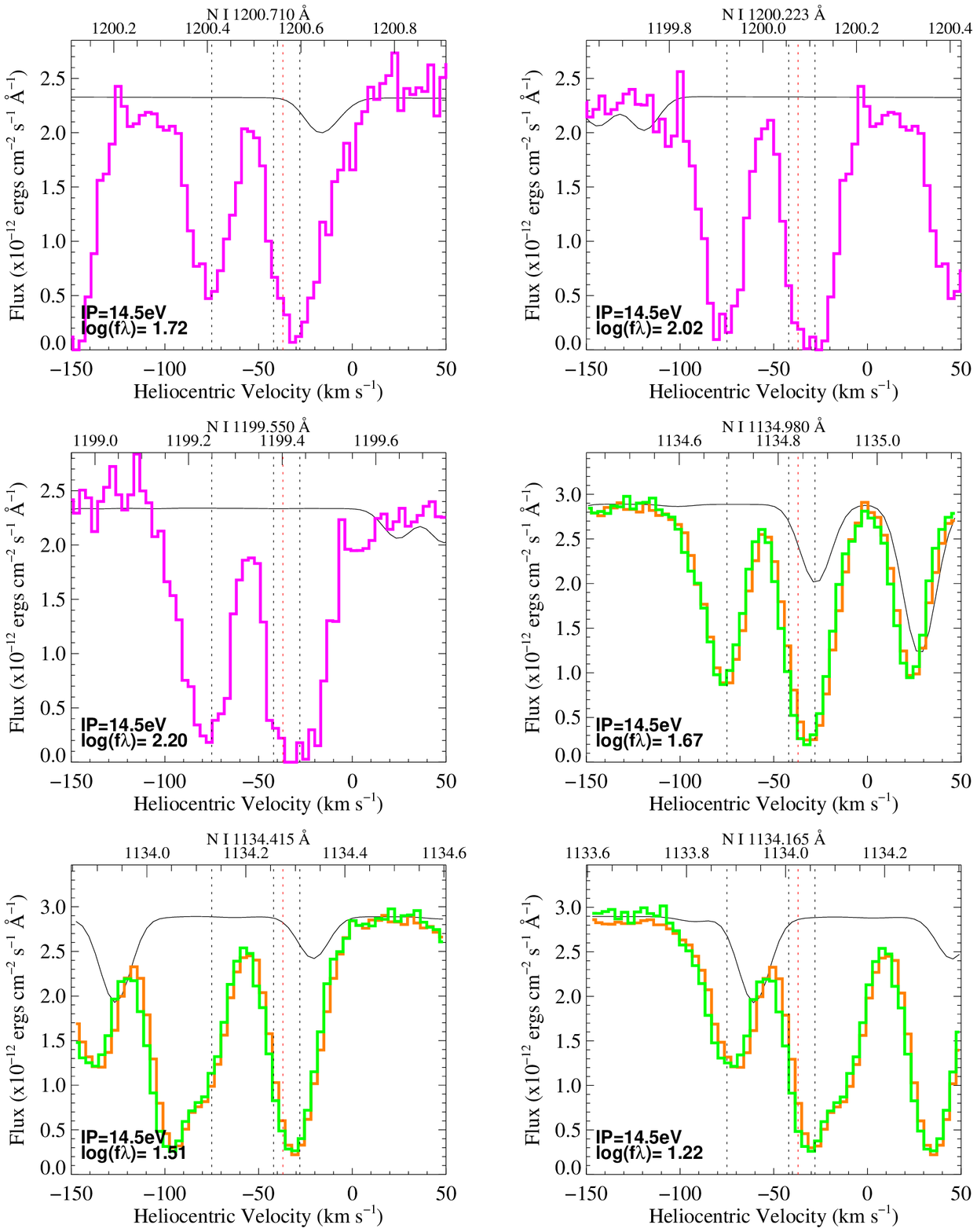}
\figcaption[f10a.eps, 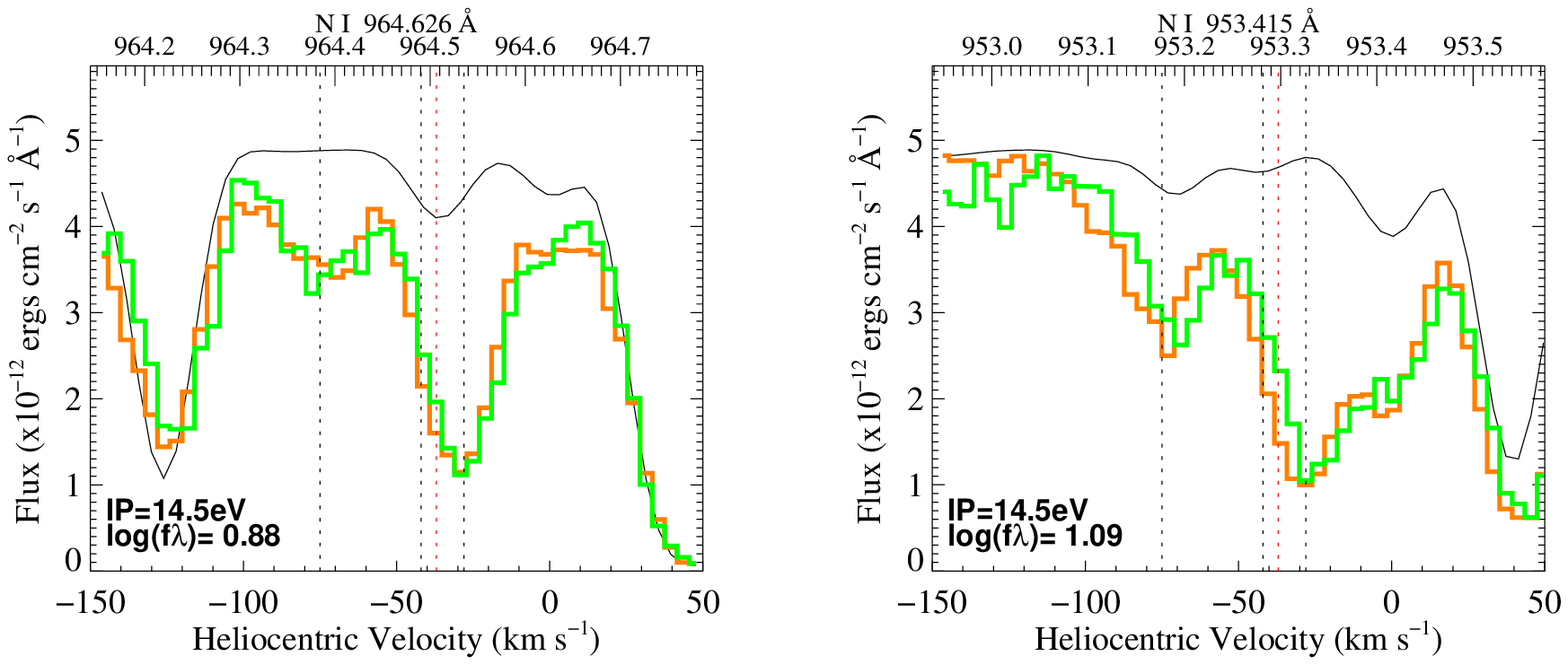]{ \label{ni0}
Absorption as a function of velocity for the \ion{N}{1} lines show
unsaturated absorption at --75 \kms.  See Figure~\ref{oviov0} for a
description of the colors.
}
\end{figure*}

\begin{figure*}[t]
\figurenum{10b}
\includegraphics[height=8.5in]{f10b.eps}
\vspace*{-5.5in}
\figcaption[]{More \ion{N}{1} lines. See Figure~\ref{oviov0} for a
description of the colors.}
\end{figure*}

\setcounter{figure}{10}

\begin{figure*}
\includegraphics[height=8.5in]{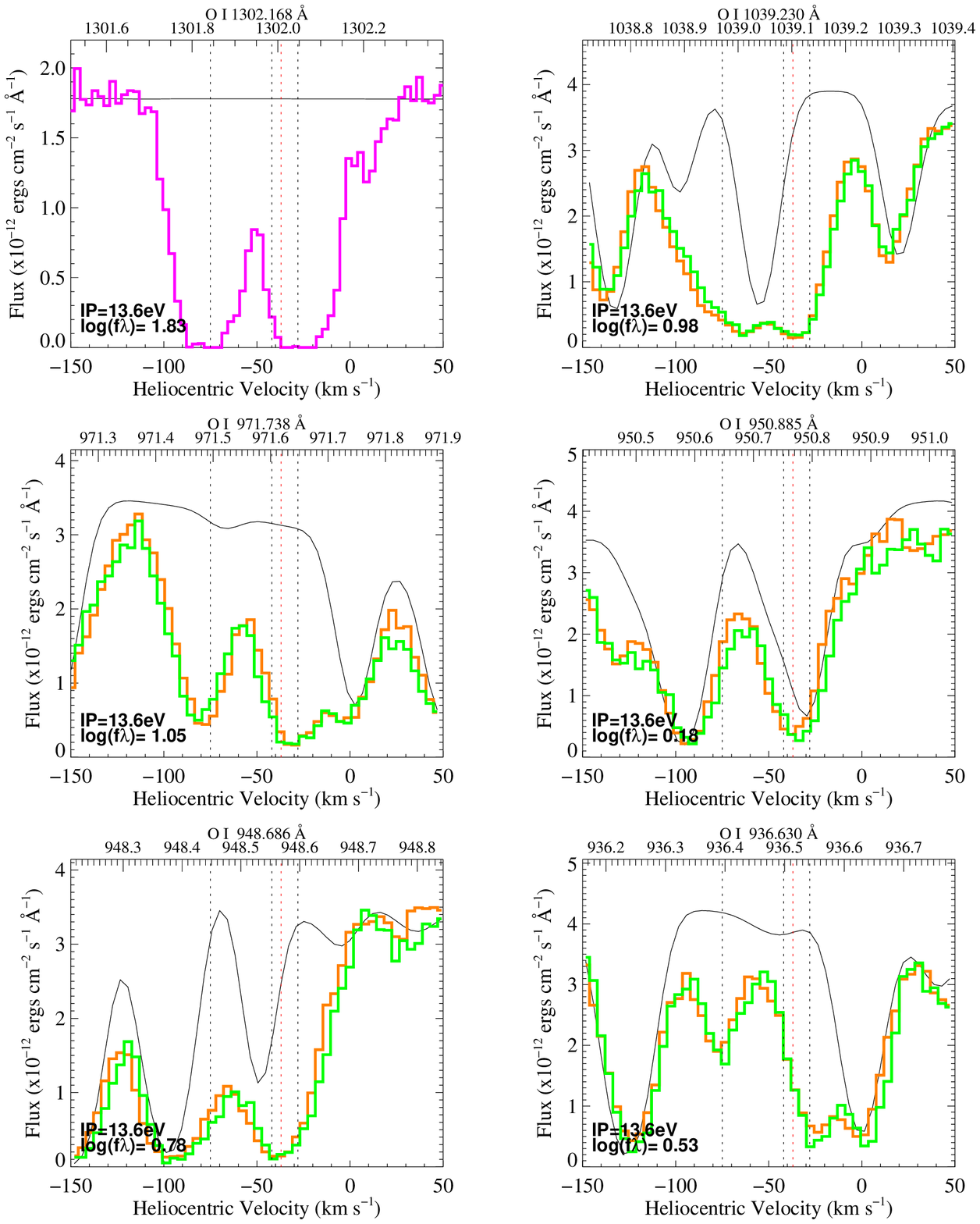}
\figcaption[f11.eps]{ \label{oi}
Absorption as a function of velocity for the \ion{O}{1} ground state
lines.  See Figure~\ref{oviov0} for a description of the colors.
} 
\end{figure*}

\begin{figure*}
\figurenum{12a}
\includegraphics[height=8.5in]{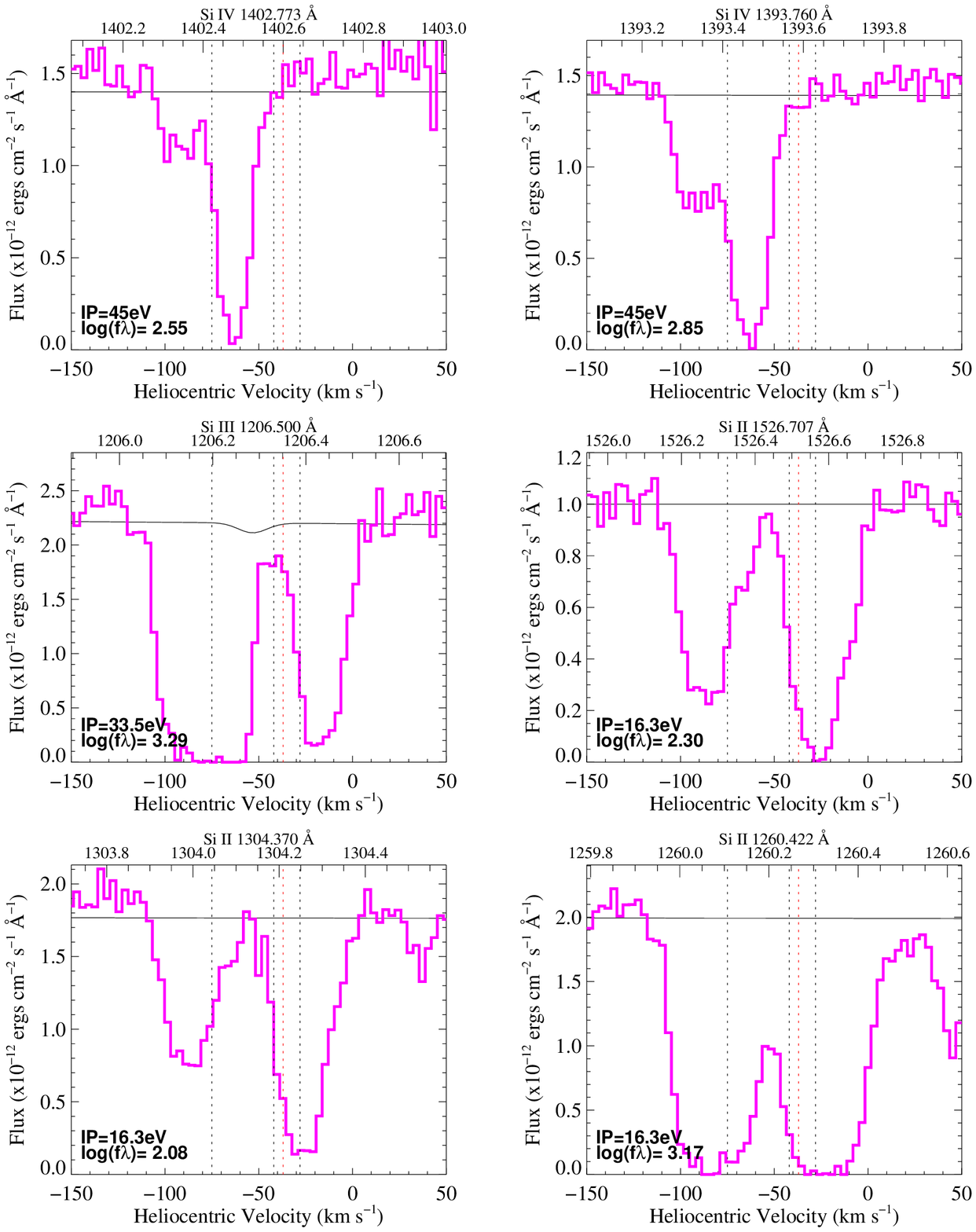}
\figcaption[f12a.eps, 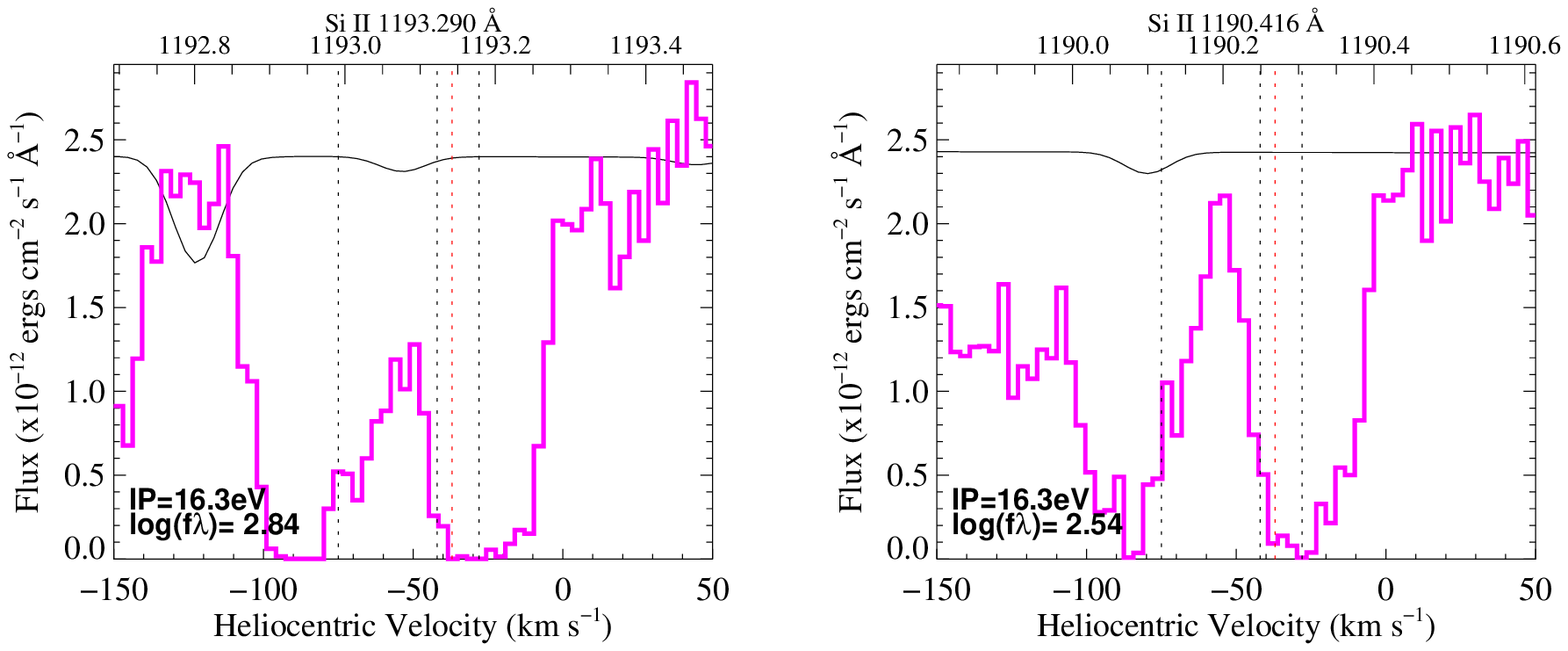]{ \label{si0}
Comparision of the \ion{Si}{4}, \ion{Si}{3} and  \ion{Si}{2}.
\ion{Si}{2}   dominates \ion{Si}{4} in the zone between --75 and --110
\kms\ while \ion{Si}{4} dominates the lower ionization species near
--60 \kms.  See Figure~\ref{oviov0} for a description of the colors.
} 
\end{figure*}

\begin{figure*}
\figurenum{12b}
\includegraphics[height=8.5in]{f12b.eps}
\vspace*{-5.65in}
\figcaption[]{More \ion{Si}{2} lines. See Figure~\ref{oviov0} for a description of the colors. }
\end{figure*}
\setcounter{figure}{12}

\begin{figure} \hspace*{-.35in}
\includegraphics[height=8.5in]{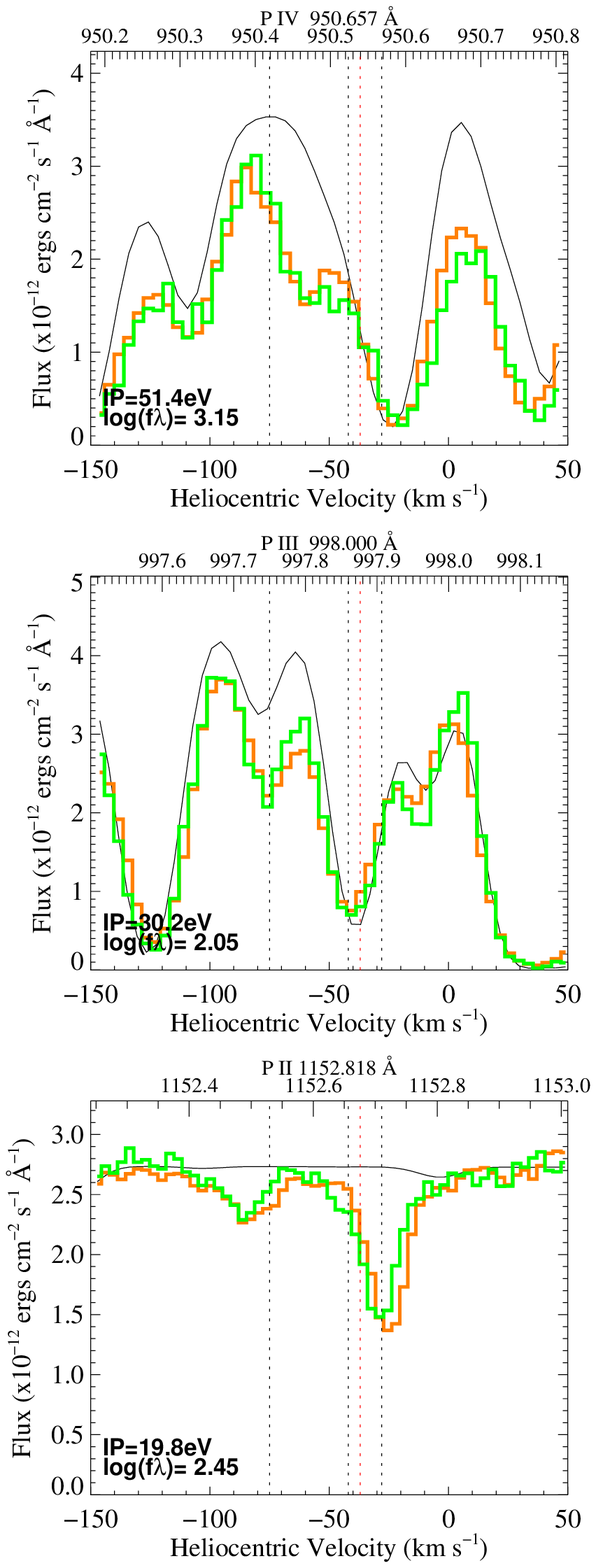}
\vspace*{.35in}
\figcaption[f13.eps]{ \label{pivpiiipii}
\ion{P}{4} \lam 950.657, \ion{P}{3} \lam 998.000, and \ion{P}{2} \lam
1152.818 lines are weak.  Although the \ion{P}{4} and \ion{P}{3} are
surrounded by molecular hydrogen they show behavior similar to Si.  See
Figure~\ref{oviov0} for a description of the colors.
}
\end{figure}

\begin{figure*}
\includegraphics[height=8.5in]{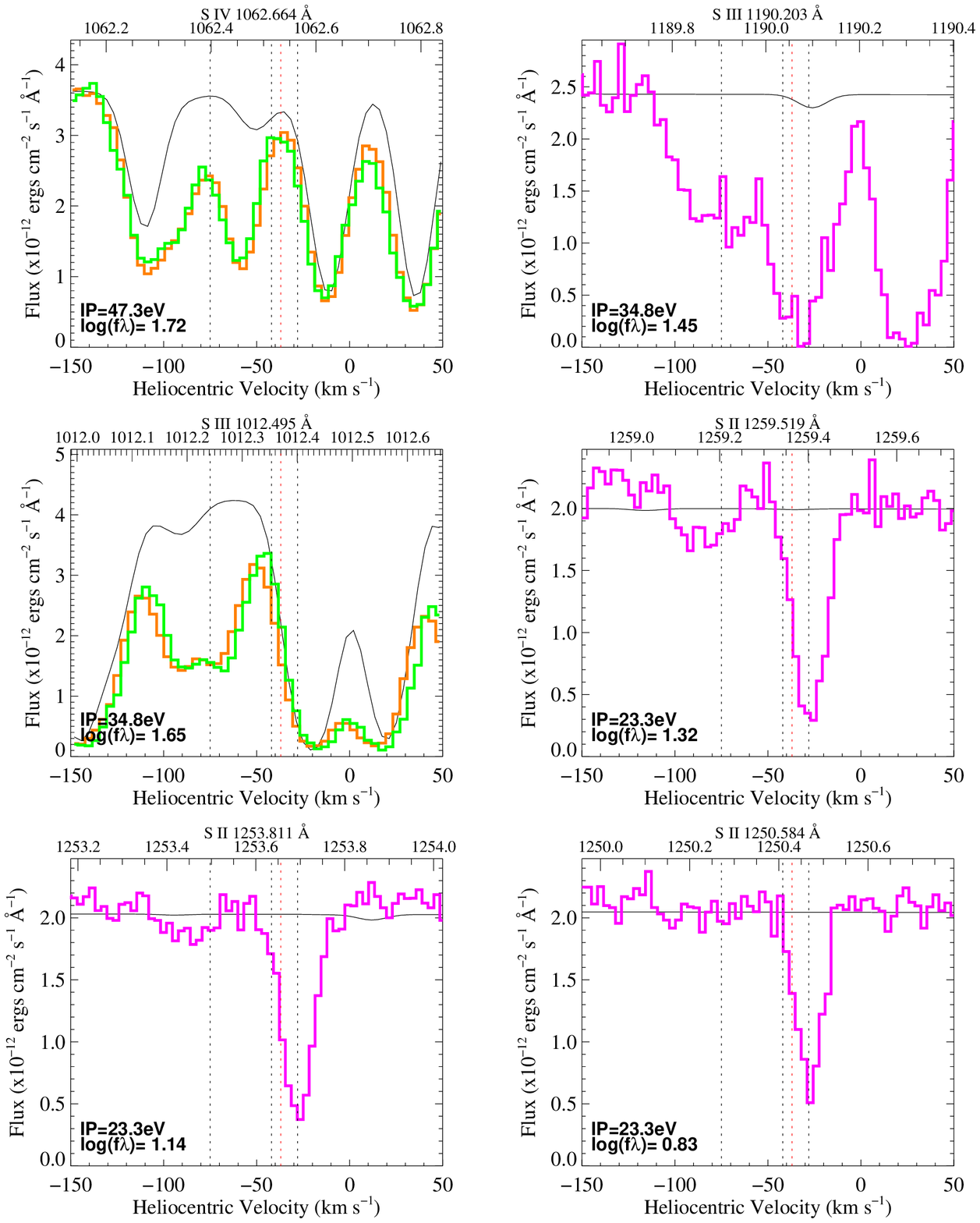}
\figcaption[f14.eps]{ \label{s}
\ion{S}{4} -- \ion{S}{2} lines show behavior similar to Si.  See
Figure~\ref{oviov0} for a description of the colors.
}
\end{figure*}

\begin{figure*}[t]
\includegraphics[height=8.5in]{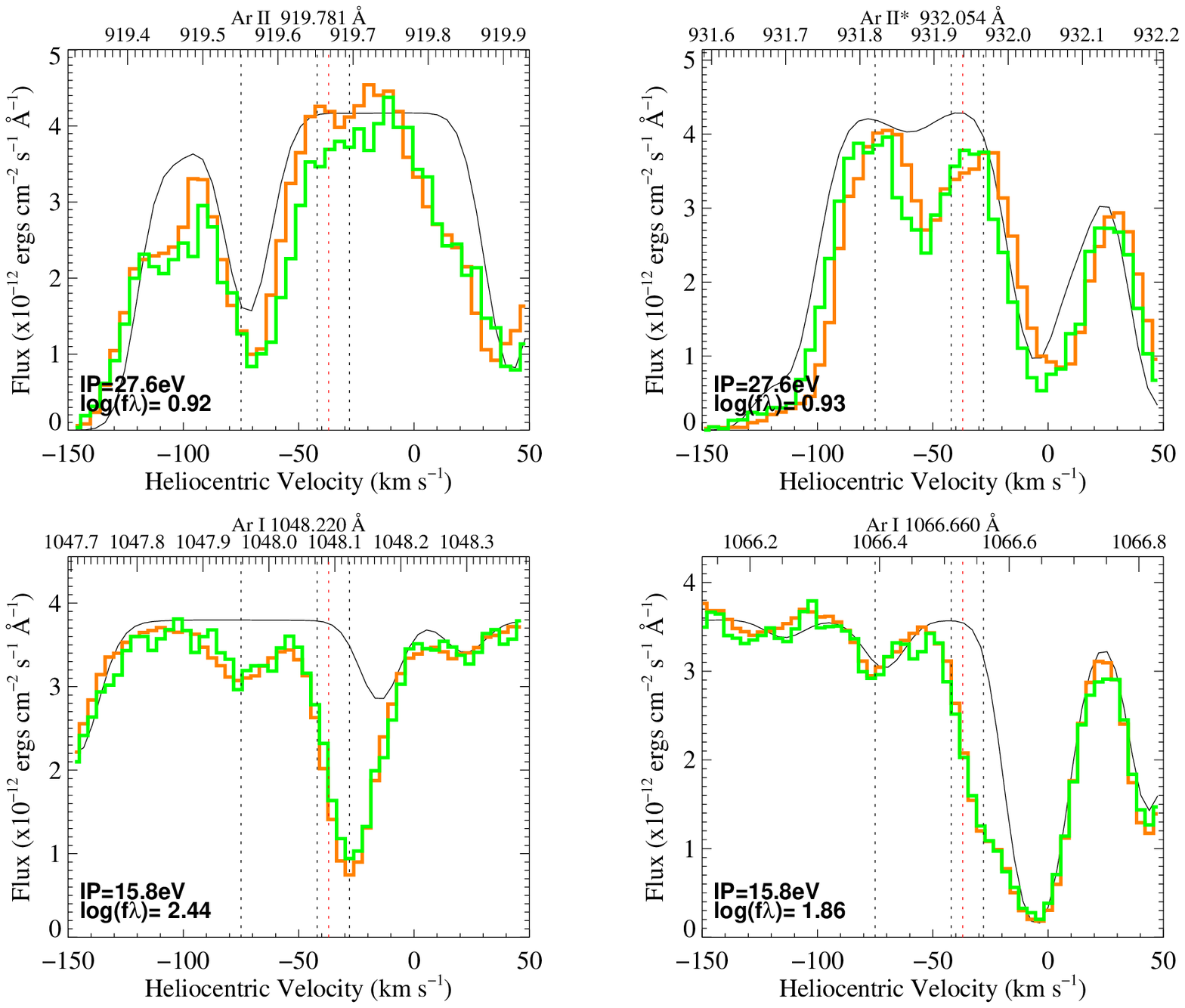}
\vspace*{-3.in}
\figcaption[f15.eps]{ \label{ar}
\ion{Ar}{2} lines appear at a velocity of $\approx$ --55 \kms, while
the \ion{Ar}{1}  appear at $\approx$ --75 \kms.  See
Figure~\ref{oviov0} for a description of the colors.
}
\end{figure*}

\begin{figure*}
\includegraphics[height=8.5in]{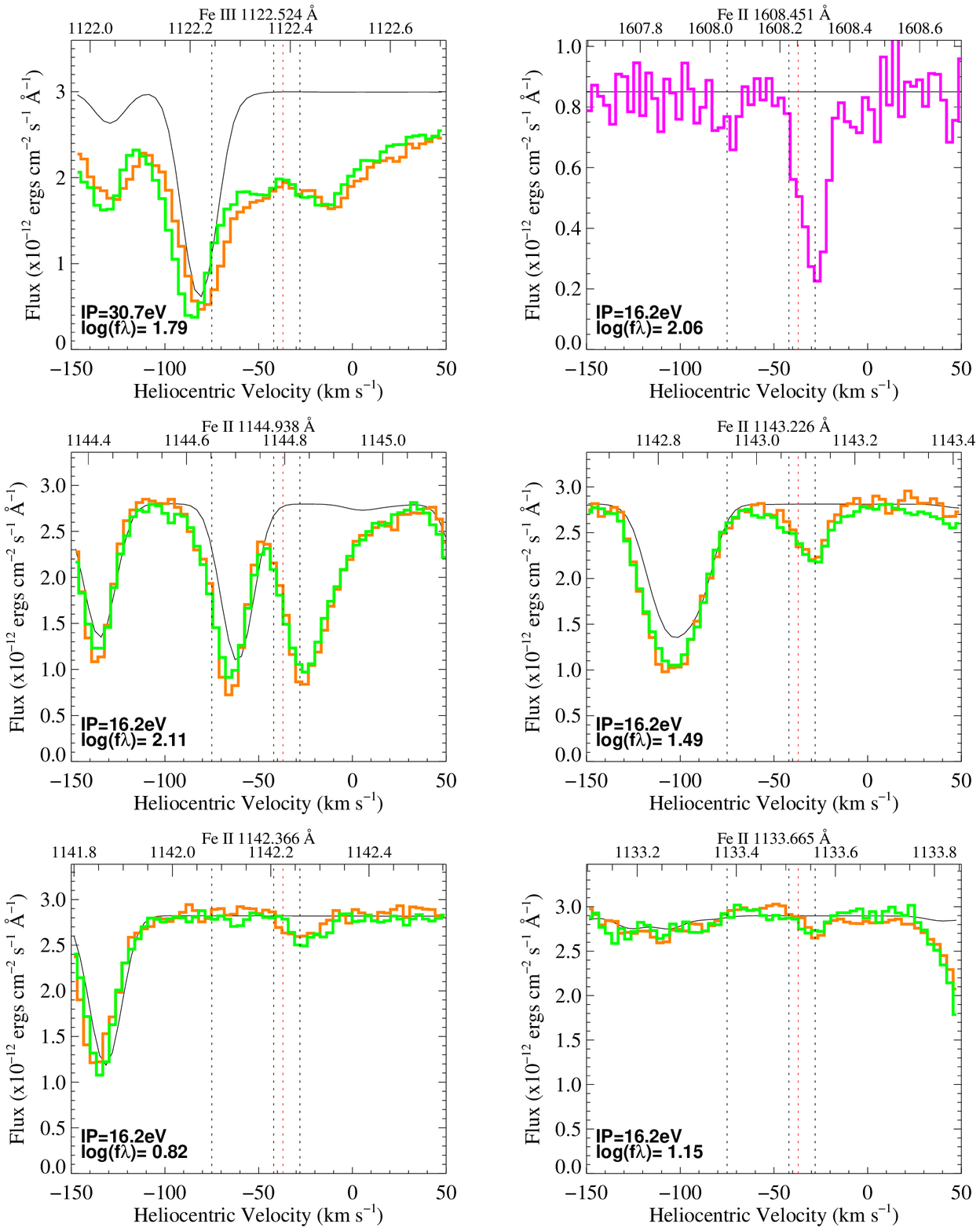}
\figcaption[f16.eps]{\label{fe}
The detection of \ion{Fe}{3} cannot be claimed.  \ion{Fe}{2} is
underabundant relative to \ion{S}{2}.  See Figure~\ref{oviov0} for a
description of the colors.
} 
\end{figure*}

\section{STIS and FUSE Absorption Profiles as a Function of Velocity }
\label{profiles}

We separate the display of individual absorption profiles in Figures~\ref{oviov0} -- \ref{ar} into
four categories, stellar, nebular plus stellar, nebular CNO and nebular
metals.   The wider stellar lines we plot over a velocity bandpass of
--250 to +150 \kms or --200 to +100 \kms.  Otherwise we use a bandpass of
--150 to +50 \kms.  We omit the broad \ion{H}{1} and \ion{He}{2}
photospheric features. The black dashed vertical lines mark the
location of the hot molecular hydrogen component at --75 \kms\, the
systemic velocity at --42 \kms, and the cold molecular hydrogen
component at --30 \kms.  The red dashed vertical lines mark the
gravitational redshift of the systemic velocity.

The s12 and s21 spectra are shown in orange and green respectively. The
continuum model, including the hydrogen lines, is plotted as a thin
black line.  Regions where the agreement is poor between the wavelength
scales for the s12, s21, and model spectra become readily apparent as do regions where the continuum flux is too high, especially
near the \ion{O}{6} lines.
For lines in the STIS region the spectra are plotted in green.

\subsection{Stellar Photospheric Features and the Absolute
Wavelength Scale}
\label{abslam}

Comparison of the \FUSE\ and STIS wavelength scales revealed
a systematic offset.  The reconciliation of this offset is essential for investigating the kinematics of the nebular outflow, where we seek to determine
the velocity of the various molecular and atomic features with respect to the systemic velocity of the nebula.
In principle, lines that arise from the photosphere should match the 
systemic velocity of the system less the gravitational redshift offset.
In practice this has been difficult to realize except for the narrowest of
photospheric features.

We have examined the overlap of the spectra in the common wavelength
regions below 1190 \AA, where high excitation photospheric \ion{C}{4}
and \ion{O}{6} lines are located.  We have also checked the
registration of the excited molecular hydrogen lines that appear in the
STIS bandpass above 1190 \AA\ with the continuum plus hydrogen model.
The disagreements revealed from these examinations have been reconciled
as described below.

An absolute reference to the heliocentric velocity was established by close examination of the \ion{O}{5} \lam 1371.296 feature.
This narrow line is a transition between two highly excited states in
\ion{O}{5} and is expected to be an excellent indicator of the
photospheric restframe (Pierre Chayer private communication).  For a
compact object of the mass and radius  found by \citet{Napiwotzki:1999} we expect
the photospheric lines to experience a gravitation redshift,
$V_{gr}=c((1-\frac{2GM_{*}}{R_{*}c^2})^{-\onehalf}-1$) = 5.1~\kms.
Applying a shift of --13 \kms\ to the STIS spectrum placed the centroid
of the \ion{O}{5} \lam 1371.296  at --37 \kms, as  expected for a
$V_{sys}$ of -42 \kms\ \citep{Wilson:1953}.

In the original analysis of the \FUSE\ M27 spectra by
\citet{McCandliss:2001a} the hot nebular molecular hydrogen component
was defined to be at --69 \kms.  Examining the overlap of common
wavelength features in the \FUSE\ and STIS spectra after defining the
\ion{O}{5} \lam 1371.296  to be at --37 \kms, we found it necessary to
shift the \FUSE\ spectra blueward by --6 \kms, such that the hot nebular
molecular hydrogen is now at --75 \kms.  The most useful overlap lines
for assessing the alignment were the doublet blend of \ion{C}{4} \dlam
1168.849, 1168.993 and the the narrow \ion{O}{6} \dlam1171.56, 1172.44.  We
note that the wavelengths of the \ion{O}{6} doublet given in the NIST online tables appear to be in error by $\approx$  --0.42
\AA.  \citet{Jahn:2006} have produced an new empirical set of \ion{O}{6} wavelengths, which agree to within 0.04 \AA\ of those found here.
We also examined the  \FUSE\ \ion{N}{1} multiplets at \dlam
1134 -- 1135 and the STIS \ion{N}{1} multiplets at \dlam 1200 -- 1201
to confirm the wavelength reconciliation.

\paragraph{\ion{O}{6} and \ion{O}{5}e} In Figure~\ref{oviov0} and ~\ref{oviov1} we show the \ion{O}{6} and
\ion{O}{5} profiles  we have identified as being photospheric in
origin.  Those lines that arise from absorption out of energy levels
well above the ground state are designated as either \ion{O}{6}e or
\ion{O}{5}e.  The  \ion{O}{6} \lam1037.62 resonance line shows some very slight signs of blue shifted nebular absorption.   

\paragraph{\ion{C}{4}e} In Figure~\ref{cive} we show the excited \ion{C}{4}e
lines.  None of these lines is saturated.   Overlapping STIS spectra if
available are plotted in purple (and grey if two orders are
available).   These spectra are comparatively noisy, but the alignment
with the \FUSE\ spectra agrees as well as the alignment between the s12
and s21 spectra.  We note that the error in the systemic velocity
is of order the STIS resoution element and is three
times the \FUSE\ resolution element.  We consider the agreement of line
profiles from spectra acquired with two different instruments to be
excellent.  The STIS order, shown in grey, of the \ion{C}{4}e \dlam1168.849, 1168.993 doublet has a spurious absorption feature that does not appear in the order shown in purple nor in the \FUSE\ spectra and can be ignored.

\subsection{Photospheric + Nebular Features}

\paragraph{\ion{C}{4} and \ion{N}{5}}
The high ionization resonance doublets \ion{C}{4} \dlam 1548.204, 1550.781,
and \ion{N}{5} \dlam 1238.821, 1242.804 show signs of nebular
absorption to the blue of --37 \kms\ and photospheric absorption to the
red as can be seen in Figure~\ref{civnv}. The nebular absorption component in the \ion{N}{5} lines is strong only between --42 and --75 \kms\ and is just barely saturated at $\approx$ -60 \kms.  In contrast, the nebular absorption component in the \ion{C}{4} lines spans
--42 to --115 \kms\ and is completely saturated from --50  to --95 \kms.

\subsection{Nebular CNO}

In general, these lines show absorption from the nebula to the blue,
and varying strengths of the intervening (non-nebular) ISM absorption
components to the red of the systemic velocity.  The nebular features
that show absorption only in the nebular expansion (i.e. at velocities
blueward  of the systemic velocity) are from intermediate or low
ionization species.    The non-nebular velocity features are most
prominent in the lowest ionization and neutral species along the
line-of-sight.

\paragraph{\ion{C}{3} and \ion{C}{2}} We show lines of 
\ion{C}{3} and \ion{C}{2} in Figure~\ref{ciiicii}.  Like the \ion{C}{4} lines, these lines are heavily saturated throughout the
nebular flow region blueward of  --42 \kms.   The saturation
makes it difficult to tell, which if any, ion is dominant in the different nebular velocity regimes.

\paragraph{\ion{N}{3}, \ion{N}{2} and \ion{N}{1}} The \ion{N}{3} lines in Figure~\ref{niii} are strongly blended with overlapping molecular hydrogen.  The blended profile indicates the absorption is saturated between --60 and --100 \kms. 
The \ion{N}{2}* and \ion{N}{2}** lines in Figure~\ref{nii} are similarily messy.  However, \ion{N}{2} \lam 1083.994
is relatively clean. It shows absorption throughout the flow, being less saturated at low velocities and becoming completely  saturated at --75 \kms.
This line also shows stonger nebular absorption than non-nebular absorption
emphasizing its dominance over \ion{N}{1}, which shows the opposite  behavior. The \ion{N}{1} multiplets at \dlam 1134.165 -- 1134.980 and  \dlam 1199.550 -- 1200.710, in Figure~\ref{ni0}, simply show unsaturated absorption centered at --75 \kms.  

\paragraph{\ion{O}{1}}  The \ion{O}{1} lines in Figure~\ref{oi} show a range of saturated to unsaturated
profiles centered at --75 \kms.  It may be possible to determine the 
column density of this species fairly accurately with a curve of growth,
after properly accounting for the continuum placement and molecular hydrogen optical depth subtraction.

\subsubsection{Nebular Metals}

The transition zone between high ionization and low ionization occurs at the
velocity of --75 \kms, where \ion{H}{1}, \ion{C}{1}, \ion{N}{1}, \ion{O}{1} and molecular hydrogen show up most strongly in the nebula.

\paragraph{\ion{Si}{4}, \ion{Si}{3} and \ion{Si}{2}}The high ionization low velocity, low ionization high velocity dichotomy
is most clearly seen in the \ion{Si}{4} \lam 1393.760 and \ion{Si}{2} \lam 1193.290 lines as shown
in Figure~\ref{si0}.  Since these lines have nearly equal $\log(f\lambda)$ they are reliable indicators of the relative ionization as a function of velocity.  \ion{Si}{2} is
stronger than \ion{Si}{4}  in the zone between --75 and --110 \kms, while
\ion{Si}{4} is stronger than \ion{Si}{2}  in the zone between --42 and --75 \kms.
The \ion{Si}{3} \lam 1206.500 line is saturated throughout most of the flow.  It is a very strong line with a $\log(f\lambda)$ = 3.29, which 
makes it difficult to determine if this species is the dominant ion throughout the flow.

\paragraph{\ion{P}{4}, \ion{P}{3} and \ion{P}{2}} We detect weak \ion{P}{4} \lam 950.657, \ion{P}{3} \lam 998.000, and \ion{P}{2}
\lam 1152.818 at progressively high velocities of  $\approx$ --65, --70, --80 \kms\ respectively as shown in Figure~\ref{pivpiiipii}. The \ion{P}{5} \lam  1117.977 line (not shown) of the \ion{P}{5} doublet is very weak, while \lam 1128.008 is  blended with \Htwo.

\paragraph{\ion{S}{4}, \ion{S}{3} and \ion{S}{2}} The \ion{S}{4} -- \ion{S}{2}
ions show similar behavior to Si and P in Figure~\ref{s}.  The \ion{S}{3} \lam1190.203 line appears to have a feature near --42 \kms.  However, this is
just the nebular component
of the nearby \ion{Si}{2} \lam 1190.416 line.  The nebular portion of this
feature is clean and agrees well with the shape of the \ion{S}{3} \lam1012.495
profile, with the absorption stongest near --70 \kms and weakening slowly to the
blue and relatively quickly to the red. 

\paragraph{\ion{Ar}{2} and \ion{Ar}{1}} Figure~\ref{ar} shows \ion{Ar}{2} \lam 919.781, \ion{Ar}{2}*  932.054, and \ion{Ar}{1}  1048.220  are detected, with \ion{Ar}{2} appearing at $\approx$ --55 \kms\
and \ion{Ar}{1} appearing at $\approx$ --75 \kms.  \ion{Ar}{1}  1066.660
is blended with an overlapping \Htwo\ line.  \ion{Ar}{2}  919.781
is blended with a nearby molecular hydrogen line but the nebular \ion{Ar}{2}* \lam 932.054 is clearly visible without model subtraction.  The thermal production of  \ion{Ar}{2}* requires a temperature $\sim$ 10,000 K, and is a confirmation that
the gas at low velocities is much hotter than that indicated by the molecular hydrogen.  The absorption of the \ion{Ar}{2} lines is stronger than  the \ion{Ar}{1} lines even though they
have transition strengths a factor of 10 lower.  This suggests the
\ion{Ar}{2} is the dominant over  \ion{Ar}{1}in the nebula.

\paragraph{\ion{Fe}{3} and \ion{Fe}{2}} 

\ion{Fe}{3} \lam 1122.524
may be blended with a molecular hydrogen line 
aon the blue side of the photospheric \ion{O}{6}e
\dlam 1122.593 -- 1122.618 doublet as shown in Figure~\ref{fe}.  
However, without a good idea of
what the photospheric line shape is in this region, we cannot claim a
detection.  The strongest transitions of
\ion{Fe}{2}  \lam 1608.451 and \lam 1144.938 appear weakly at --75
\kms.  These lines have transition strengths ($\log{(f\lambda)}$ =
2.06, 2.11) that are nearly identical to that of \ion{Si}{2}
\lam 1304.370 ($\log{(f\lambda)}$ = 2.08), yet the \ion{Si}{2} feature
is much stronger. The ionization potential of \ion{Si}{2} (16.3 eV) is
also nearly identical to that of \ion{Fe}{2} (16.2 eV) and they also
have similar condensation temperatures \citep{Morton:2003}.  We
conclude that \ion{Fe}{2} relative to \ion{Si}{2} is underabundant in
the nebula.

Arguing in a similar vein, we see that the transition at \ion{Fe}{2}
\lam 1133.665 is not detected in the nebula at all.  It has nearly the
same transition strength ($\log{(f\lambda)}$ = 1.15) as the \ion{S}{2}
\lam 1253.811 line ($\log{(f\lambda)}$ = 1.14), which is weak but
easily seen.  We conclude that \ion{Fe}{2}  relative to \ion{S}{2} is
also underabundant.  The solar abundance of Si and Fe are nearly
identical (7.56, 7.50 respectively) and are slightly higher than S
(7.2).  It is not unusual to have the Si abundance greater than the Fe
in the typical ISM because it is depleted onto dust \citep{Savage:1996}.
However, here we have no Fe in the gas phase and appearently very little dust
in the diffuse nebular medium.  This is a puzzle.

\section{Suggestions for Future Investigations}

M27 is a excellent laboratory for testing theories regarding the
abundance kinematics in PNe because the stellar temperature, mass,
gravity, and the nebular mass, distance, abundances, extinction and
excitation states, are well quantified.  The 
reduced spectra provided by this study and the associated atomic
and molecular hydrogen model for the nebular and non-nebular
absorptions, should enable a photospheric analysis of the metal abundances, similar to the effort undertaken by \citet{Jahn:2006} for PG 1159 - 035.  Establishing a reliable stellar continuum is
the first step towards determining accurate nebular absorption line
abundances for comparision with  those derived from the emission line analyses
of \citet{Barker:1984} and \citet{Hawley:1978}.  

The finding in Paper I, of little extinction by dust in the nebula 
raises a number of interesting questions concerning the depletion of
metals in the diffuse and clumpy media, which go beyond the scope of
this investigation.  A major limitation to this effort is the
requirement for a good stellar model that can reliably reproduce the
observed photospheric features and continuum, especially in the
vicinity of the blend of  nebular \ion{Fe}{3} \lam 1122.52  with
photospheric \ion{O}{6} \lam 1122.62.  If such a stellar model can be
produced, the total S, Si, P and Fe abundances should be derivable by
profile modeling after accounting for the atomic and molecular hydrogen absorption provided here.  The question of whether the abundances of these metals change across the neutral transition zone could provide information on
whether photo-evaporation of the globules is an important process for the
enrichment of metals in the high velocity zone.  A model of the nebular
ionization as a function of velocity would be useful for this purpose.
Such a model might be produced by melding the detailed time dependent approach to the calculations of the atomic and molecular emissions followed by  \citet{Natta:1998}, with the radiative hydrodynamical rigor of \citet{Villaver:2002b}.
The problem may also be approached by using the Sobolev plus exact
integration (SEI) method  of \citet{Lamers:1987}, from which a empirical
parameterization of the wind velocity `law'' could be obtained.

\acknowledgments

We are grateful to Patrick J. Huggins who provided encouragement to
complete this work.  Thanks also to Roxana Lupu, Kevin France and Eric
Burgh for stimulating suggestions. Based on observations made with the
NASA-CNES-CSA Far Ultraviolet Spectroscopic Explorer. \FUSE\ is
operated for NASA by the Johns Hopkins University under NASA contract
NAS5-32985.  In addition, some of the data presented in this paper were
obtained from the Multimission Archive at the Space Telescope Science
Institute (MAST). STScI is operated by the Association of Universities
for Research in Astronomy, Inc., under NASA contract NAS5-26555.
Support for MAST for non-HST data is provided by the NASA Office of
Space Science via grant NAG5-7584 and by other grants and contracts.



Facilities: \facility{HST} (STIS), \facility{FUSE}.






;\clearpage




\end{document}